\documentstyle[newhutp]{article}
\makeatletter \input art10.sty \makeatother
\def\starttext{\twocolumn}

\typein{*** or hit [return] for landscape mode(11x8.5) ***}
\newcommand{\grotz}{h}
\newcommand{\Grotz}{H}
\newcommand{\beq}{\begin{equation}}
\newcommand{\eeq}{\end{equation}}

\newcommand{\remove}[1]{}
\renewcommand{\theequation}{\thesection.\arabic{equation}}

\newdimen\pmboffset
\def\oldpmb#1{\setbox0=\hbox{#1}%
 \copy0\kern-\wd0
 \kern\pmboffset\raise 1.732\pmboffset\copy0\kern-\wd0
 \kern\pmboffset\box0}

\begin{document}

\title{Technicolor with a Massless Scalar Doublet\thanks{Research supported in
part by the National Science Foundation,
under grant \# PHY-9218167, and in part by the Texas National
Research Laboratory Commission under grant \# RGFY93-278B.}}

\author{
    Christopher D. Carone\thanks {
   e-mail address: carone@huhepl.harvard.edu
 } \\ Howard Georgi\thanks { e-mail address: georgi@huhepl.harvard.edu } \\
    Lyman Laboratory of Physics \\
    Harvard University \\
    Cambridge, MA 02138
}
\date{\today}

\renewcommand{\preprintno}{HUTP-93/A015}

\begin{titlepage}

\maketitle

\def\thepage {}    

\begin{abstract}
We consider a minimal technicolor model in which the ordinary
and technicolor sectors are coupled by a {\it massless} scalar
doublet. When technicolor interactions become strong, the
resulting technicolor condensate not only breaks the electroweak
symmetry, but also causes the scalar to develop a vacuum expectation value.
With the appropriate choice of the scalar's Yukawa couplings, fermion
masses are generated, giving us the conventional pattern of flavor
symmetry breaking. Although no explicit scalar mass term appears in the
full lagrangian of the model, the pseudoscalar states that remain in the
low-energy effective theory gain sufficient mass through technicolor
interactions to evade detection. We show that this model does not
generate unacceptably large flavor changing neutral currents, and
is consistent with the experimental constraints on oblique electroweak
radiative corrections. We determine the experimentally allowed region
of the model's parameter space, and discuss the significance of a
phenomenologically viable model that has no arbitrary dimensionful
parameters.  In terms of parameter counting, our model is the simplest
possible extension of the standard model.
\end{abstract}

\end{titlepage}

\starttext 
\pagestyle{columns} 
\pagenumbering {arabic} 

\typeout{**************************************************************}
\typeout{*** Producing the figures in PICTEX takes time and memory. ***}
\typeout{*** You do not have to do it on the first LATEX pass. If   ***}
\typeout{*** you have trouble compiling the figures, just skip it.  ***}
\typeout{**************************************************************}
\typein[\yorn]{*** Enter `y' to compile the figures -- `n' to skip ***}
\if y\yorn
\input prepictex
\input pictex
\input postpictex
\newdimen\tdim
\tdim=.14ex
\def\stpltsmbl{\setplotsymbol ({\footnotesize .})}
\newbox\phru
\setbox\phru=\hbox{\beginpicture
\setcoordinatesystem units <\tdim,\tdim>
\stpltsmbl
\setquadratic
\plot
0 0
2.5 3
5 0
7.5 -3
10 0
/
\endpicture}
\def\photonru #1 #2 *#3 /{\multiput {\copy\phru}  at
#1 #2 *#3 10 0 /}
\newbox\phdr
\setbox\phdr=\hbox{\beginpicture
\setcoordinatesystem units <\tdim,\tdim>
\stpltsmbl
\setquadratic
\plot
0 0
3 -2.5
0 -5
-3 -7.5
0 -10
/
\endpicture}
\def\photondr #1 #2 *#3 /{\multiput {\copy\phdr}  at
#1 #2 *#3 0 -10 /}
\def\tarrow{\arrow <5\tdim> [.3,.6]}
\def\finput#1{#1}
\def\fnewpage{\newpage}
\else
\def\finput#1{\par}
\def\fnewpage{}
\fi
\setcounter{bottomnumber}{2}
\setcounter{topnumber}{3}
\setcounter{totalnumber}{4}
\newcommand{\figsize}{\small}
\renewcommand{\bottomfraction}{1}
\renewcommand{\topfraction}{1}
\renewcommand{\textfraction}{0}

\section {Introduction} \label {sec:intro}

Technicolor models provide an elegant mechanism for electroweak
symmetry breaking, but give no natural explanation for
the generation of fermion masses. A number of different scenarios
have been proposed to solve this problem, including extended
technicolor (ETC) models~\cite{etc}, and composite technicolor standard
models (CTSM)~\cite{ctsm}, but neither of these has
been free of other serious shortcomings. ETC models that can
accommodate a heavy top quark often can do so only at the expense of
generating large flavor changing neutral current (FCNC) effects.
While CTSM models have a GIM mechanism built into the
technicolor sector to minimize this problem, all the realistic examples
that have been proposed require many new gauge groups beyond those of
the standard model, and therefore these models become aesthetically
unappealing. A simple alternative to ETC and CTSM models that suffers from
neither of these drawbacks is technicolor with a scalar
doublet~\cite{bostech}. The scalar communicates electroweak symmetry breaking
to the ordinary fermions in a way that does not generate large FCNC effects.

In previous phenomenological studies of technicolor with a single
scalar doublet, it was assumed that the scalar had a significant
$SU(2)\times U(1)$ invariant mass, and that quartic terms in the scalar
potential could be ignored. The resulting analysis established that
the single scalar doublet model was phenomenologically viable over a
wide range of the model's parameters. In particular, the model could
account for a heavy top quark without generating large flavor changing
neutral currents, and without exceeding the experimental bounds
on the electroweak $S$ and $T$ parameters~\cite{carone}. Part of
this success was achieved at the expense of allowing an undetermined
dimensionful parameter in the theory, namely, the mass of the scalar
doublet, $M_\phi$. A simple way to account for this unknown scale is
to assume that the scalar is composite, and that it's mass is calculable
given knowledge of the detailed dynamics of a full, high-energy theory.
In this paper, we will explore another alternative, namely, that the scalar
is fundamental and that $M_\phi\equiv 0$. In this limit, the lagrangian
for our model contains no arbitrary dimensionful parameters. Although the
original scalar doublet is now massless, we will see that technicolor
interactions are sufficient to give the physical scalar states in the
low-energy effective theory masses large enough to insure that they are
not detected. In addition, we will show that FCNC effects are not
unacceptably large, and electroweak oblique corrections do not
exceed current experimental bounds. In short, the model is
again phenomenologically viable. Furthermore, because of the absence of the
scalar mass term, the model has only two more parameters than the simplest
version of standard model. In this sense, it is the simplest possible
extension. In addition, in some allowed regions of the parameter space, the
model differs significantly from the more conventional technicolor models
discussed in~\cite{carone}.

The paper is organized as follows. In Section~\ref{sec:model},
we describe the basic features of the model. In Section~\ref{sec:chiral},
we construct a low-energy effective chiral lagrangian to describe the
physics of the scalar states below the electroweak scale. In Section
\ref{sec:cw}, we study the effects of Coleman-Weinberg radiative corrections
on the scalar potential. In
Section~\ref{sec:fcnc}, we set up the analysis of flavor changing
neutral currents in the model. In Section~\ref{sec:phenom}, we
determine the region in the model's parameter space that is excluded
by experimental constraints, and discuss possible experimental signatures.
In Section~\ref{sec:sandt} we discuss the oblique electroweak radiative
corrections and present our estimate of the low-energy contributions to
the $S$ and $T$ parameters. In Section~\ref{sec:beyond} we consider the
phenomenology of the region of the model's parameter space that is not
adequately described by an effective chiral lagrangian. In the final section,
we discuss the theoretical significance of a viable model that contains no
arbitrary dimensionful parameters, and we summarize our conclusions.

\section{The Model}\label{sec:model}

The gauge structure of the model is simply the direct product of
of the technicolor and standard model gauge groups:
$SU(N)_{TC}\times SU(3)_C \times SU(2)_W \times U(1)_Y$.\footnote{We will
assume that $N=4$ in all the quantitative estimates below.}
The technicolor singlet fermions are exactly those
of the standard model, in the usual $SU(2)_W$ representations
of left-handed doublets and right-handed singlets:
\begin{eqnarray}
L_{L} & = & {l \choose \nu}_{L} , \ \ \ \ \ l_{R}\,, \nonumber \\
Q_{L} & = & {U \choose D}_{L} ,\ \ \ \ U_{R},\ \ D_{R}\,.
\end{eqnarray}
Here $l \equiv (e, \mu, \tau),\ \nu \equiv (\nu_e, \nu_\mu,
\nu_\tau),\ U \equiv (u, c, t),$ and $D \equiv (d, s, b)$.
We assume that the technicolor sector is minimal, i.e. it
consists of two techniflavors, $p$ and $m$, that also transform
under $SU(2)_W$ as a left-handed doublet and two right-handed
singlets
\begin{equation}
\Upsilon_L = {p \choose m}_L,\ \ \ \ p_R,\ \ m_R\,.
\end{equation}
We assign the hypercharges $Y(\Upsilon_L)$=0, $Y(p_R)$=${1 \over 2}$,
and $Y(m_R)$=$-{1 \over 2}$ so that the model is free of gauge
anomalies. In addition, $p$ and $m$ each transform in the fundamental
representation of $SU(N)_{TC}$.

When technicolor becomes strong at a scale $\approx 4 \pi f$, the
technifermions' chiral symmetries spontaneously break, and the
technifermions form a condensate
\begin{equation}
\langle{\bar p p + \bar m m}\rangle \approx 4 \pi f^3,
\end{equation}
where $f$ is the technipion decay constant. The condensate breaks
the original $SU(2)_W \times U(1)_Y$ electroweak symmetry down
to $U(1)_{EM}$, giving mass to the W and Z bosons. The ordinary
fermions, however, are left unaffected by electroweak symmetry breaking,
and will remain massless unless we provide some additional mechanism.

To couple the ordinary fermions to the technicolor condensate, we introduce
a massless scalar field $\phi$, that transforms as an $SU(2)_W$ doublet, with
hypercharge $Y(\phi)={1 \over 2}$. The scalar has Yukawa couplings to
both the technifermions
\begin{equation}
{\cal L}_{\phi T} = \bar\Upsilon_L \tilde\phi\thinspace \grotz_+ p_R\ +\
  \bar\Upsilon_L \phi\thinspace \grotz_- m_R\ +\ \mbox{h.c.}
\label{eeq:ytc}\end{equation}
and to the ordinary fermions
\begin{equation}
{\cal L}_{\phi f} = \bar L_{L} \tilde\phi\thinspace \grotz_{l} l_{R}
 + \bar Q_{L} \tilde\phi\thinspace \grotz_{U} U_{R}\ +\
   \bar Q_{L} \phi\thinspace \grotz_{D} D_{R}\ +\ \mbox{h.c..}
\label{eeq:yordf}
\end{equation}
When the technifermions condense, the scalar develops a vacuum expectation
value that generates mass terms for the ordinary fermions. We will
see how this works explicitly in Section~\ref{sec:chiral}. The coupling
matrices $\grotz_{f}$ are proportional to the fermion mass matrices and
generate the usual pattern of flavor symmetry breaking of the standard
model; in particular, the quarks mix via the conventional CKM matrix.

The new free parameters in our model that are associated with the scalar
are the Yukawa couplings, $(\grotz_+,\grotz_-)$, and the technicolor scale,
$\Lambda_{TC}$. The value of $\Lambda_{TC}$ will be determined by the
$SU(2)\times U(1)$ breaking scale (although in a nontrivial way), just as this
scale determines the value of the scalar mass term in the simplest standard
model. Thus the two ``new'' parameters can be taken to be $\grotz_\pm$. As in
the simplest standard model, the physics also depends on the unknown
$(\phi^\dagger\phi)^2/2$ coupling in the scalar potential, which
we will call $\lambda$. Our aim is to maximally constrain
these parameters given the current experimental limits on
the relevant physical processes. However, it will be more convenient
for us to express our results in terms of the equivalent set of
parameters $\lambda$, $\grotz$, and $\delta$, where
\beq
\grotz = (\grotz_++\grotz_-)/2\,,\quad
\delta= (\grotz_+-\grotz_-)/(\grotz_++\grotz_-)\,.
\label{eq:grotz}
\eeq
In this parametrization, the technicolor sector of the model
is custodial isospin-conserving when $\delta=0$, and maximally
isospin-violating when $\delta=1$. Since custodial isospin-violation
enters at order $\delta^2\grotz^2$ in the chiral expansion, the lowest order
results we present in Sections~\ref{sec:chiral}, \ref{sec:fcnc},
and \ref{sec:phenom}, will depend on $\lambda$ and
$\grotz$ exclusively. Thus, we will devote much of our effort to
identifying the areas of the $\lambda$-$\grotz$ plane that are
excluded by experiment. The parameter $\delta$ will be of relevance in
Section~\ref{sec:sandt} where we estimate the non-standard contributions to
the $T$ parameter. We will show that $\delta$ can be set to its
maximum value throughout much of the allowed region of the
$\lambda$-$\grotz$ plane, without generating dangerously large corrections
to $T$. Thus, we will find that it is not necessary to fine-tune $\delta$
in order to satisfy the experimental constraints.

\section{The Effective Chiral Lagrangian}\label {sec:chiral}

In this section, we construct an effective chiral lagrangian for our model
to describe the physical scalar degrees of freedom below the technicolor
scale.  This approach is possible because the technicolor kinetic terms
have an $SU(2)_L \times SU(2)_R$ chiral symmetry that is broken
spontaneously to a diagonal $SU(2)_c$; the latter is the well
known custodial $SU(2)$ which prevents the $T$ parameter from deviating
greatly from zero. The pseudo-goldstone bosons that result from the
chiral symmetry breaking are the isotriplet of technipions
\beq
\Pi = \frac{1}{\sqrt{2}}\left[
\begin{array}{cc} \pi^0/\sqrt{2} & \pi^+ \\
          \pi^-     & -\pi^0/\sqrt{2}
\end{array}\right]\,.
\label{eq:pimatrix}
\eeq
The chiral lagrangian analysis would parallel that of
QCD except that the chiral symmetry of interest to us here is gauged. We
can imbed the conventional weak $SU(2)$ completely in $SU(2)_L$, and let
hypercharge be generated by the $\tau_3$ component of $SU(2)_R$.
(This imbedding is possible because the left-handed technifermion doublet
has zero hypercharge.)  We adopt the conventional nonlinear representation
of the technipion fields
\beq
\Sigma = \exp\left(2 \, i \, \Pi/f\right)\,, \quad\quad
\Sigma \rightarrow L\Sigma R^\dagger\,.
\label{eq:Sigma}
\eeq
with $\Pi$ defined in (\ref{eq:pimatrix}) and we write the
scalar doublet $\phi$ in matrix form
\beq
\Phi=\left[\begin{array}{cc} \overline{\phi^0} & \phi^+ \\
                   -\phi^- & \phi^0
\end{array}\right]\,.
\label{eq:matphi}
\eeq
Then, the kinetic energy terms are given by
\beq
{\cal L}_{{\rm K.E.}}
 = \frac{1}{2}{\rm Tr} (D_\mu \Phi ^\dagger D^\mu \Phi)+
\frac{f^2}{4} {\rm Tr} (D_\mu \Sigma ^\dagger D^\mu \Sigma)
\label{eq:ke1}
\eeq
with the covariant derivative defined by
\beq
D^\mu \Sigma = \partial^\mu \Sigma - i g W^\mu_a \frac{\tau^a}{2} \Sigma
+ i g' B^\mu \Sigma \frac{\tau^3}{2}\,.
\label{eq:coderiv}
\eeq
Following the analysis of ref.~\cite{carone}, we use the fact that
$\Phi^\dagger \Phi \propto 1$ to rewrite $\Phi$ in terms of an
isosinglet scalar field $\sigma$, and a unitary matrix $\Sigma '$:
\beq
\Phi=\frac{(\sigma+f')}{\sqrt{2}} \Sigma '
\label{eq:redef1}
\eeq
where
\beq
\Sigma ' = \exp \left(2 i \Pi'/f'\right)\,.
\label{eq:sigmap}
\eeq
Later, we will present a more elegant redefinition of $\Phi$ that
is manifestly nonsingular in the limit $f'\rightarrow 0$. Since the
physics is independent of the particular nonlinear field redefinition
we adopt, we will work for the moment with (\ref{eq:redef1})
to derive our main results. Under this redefinition, the kinetic terms
become
\beq
{\cal L}_{{\rm K.E.}}=\frac{1}{2}\partial_\mu\sigma\partial^\mu\sigma
+\frac{f^2}{4}{\rm Tr}(D_\mu \Sigma^\dagger D^\mu \Sigma)
+\frac{(\sigma+f')^2}{4}{\rm Tr}(D_\mu {\Sigma'}^\dagger D^\mu \Sigma')\,.
\label{eq:fullke}
\eeq
By expanding eq. (\ref{eq:fullke}) in terms of the component fields, it
is easy to show that the pions in the linear combination
\beq
\pi_a = \frac{f\Pi+f'\Pi'}{\sqrt{f^2+f'^2}}
\eeq
become the longitudinal components of the weak gauge bosons in
unitary gauge, while those in the orthogonal linear combination,
\beq
\pi_p = \frac{-f'\Pi+f\Pi'}{\sqrt{f^2+f'^2}}
\eeq
remain as physical scalars in the low-energy theory. In addition,
we obtain the correct gauge boson masses only if
\beq
f^2+f'^2 = \frac{\sin^2\theta \cos^2\theta}{\pi \alpha} M_Z^2\equiv v^2\,.
\label{eq:vdef}
\eeq
{}From now on we will work in unitary gauge, where the particle spectrum
consists of $\pi_p$, $\sigma$, and the massive weak gauge bosons.

Since we are interested in the determining the masses of the fields
$\pi_p$ and $\sigma$ we need to study the scalar potential. At
lowest order, it is given by
\beq
V_\lambda(\sigma)=\frac{\lambda}{8}\left[{\rm Tr}(\Phi^\dagger \Phi)\right]^2
=
\frac{\lambda}{8}\,(\sigma + f')^4\,.
\label{eq:scalpot0}
\eeq
This potential for $\Phi$ is the only one consistent with
renormalizability, gauge invariance, and our assumption that $M_\phi=0$.
At one-loop, there are additional terms in the potential of the form
$\sigma^4\log(\sigma^2/\mu^2)$
induced by radiative corrections {\it a la} Coleman-Weinberg~\cite{colwein}.
These terms are important in some regions of the parameter space, however, we
will begin by ignoring them. We hope that this will make the subsequent
analysis easier to understand.

On the other hand, we must always include the contributions to the potential
generated by the technicolor interactions. To write down all appropriate
terms consistent with the chiral symmetry, we recall that the coupling of
the scalar to the technifermion doublet $\Upsilon$ is given by
\beq
\overline{\Upsilon}_L
\left(\begin{array}{cc} \overline{\phi^0} & \phi^+ \\
                   -\phi^- & \phi^0
\end{array}\right)
\left(\begin{array}{cc} \grotz_+ & 0 \\ 0 & \grotz_-
\end{array}\right) \Upsilon_R \equiv
\overline{\Upsilon}_L \Phi \Grotz
\Upsilon_R
\eeq
where $\Grotz$ is the technifermion Yukawa coupling matrix.
Thus, if we treat the matrix $\Phi\Grotz$ as a spurion transforming
as
\beq
(\Phi\Grotz)\rightarrow L (\Phi\Grotz) R^\dagger
\label{eq:spurion}
\eeq
and build all possible invariants, we will obtain the correct
effective lagrangian. In fact, we will only require
the simplest term,
\beq
{\cal L}_\Grotz = c_1 \cdot 4\pi f^3 {\rm Tr}
(\Phi\Grotz\Sigma^\dagger) + \mbox{h.c.}
\label{eq:c1term}
\eeq
where the coefficient $c_1$ is of order unity, by
naive dimensional analysis (NDA)~\cite{manohar}. This interaction
generates a linear term in $\sigma$
\beq
V_\Grotz(\sigma)=-\sqrt{2} c_1 \cdot 4\pi f^3 (\grotz_+ +
\grotz_-) \sigma\,.
\label{eq:linterm}
\eeq
We assume that the $\sigma$ field has no vacuum expectation value, and
therefore we require that the linear terms in $\sigma$ vanish:
\beq
V'_\lambda(0)+V'_\Grotz(0)=0\,,
\label{eq:vanish}
\eeq
or
\beq
\frac{\lambda}{2}{f'}^3=8\sqrt{2} c_1 \pi\grotz f^3\,.
\label{eq:vanish1}
\eeq
Together with the constraint imposed by eq. (\ref{eq:vdef}), this
completely determines the pion decay constants, $f$ and $f'$,
in terms of the model's free parameters
\beq
f= v \left[1+(16\sqrt{2}\pi c_1)^\frac{2}{3}
(\grotz/\lambda)^{\frac{2}{3}}\right]
^{-\frac{1}{2}}\,,
\label{eq:f}
\eeq
\beq
f'= f (16\sqrt{2}\pi c_1)^\frac{1}{3}\left(\frac{\grotz}{\lambda}\right)
^\frac{1}{3}\,.
\label{eq:fprime}
\eeq
where $\grotz\equiv (\grotz_++\grotz_-)/2$. Now it is simple
to determine the scalar masses. The lowest order contribution to
$m^2_\sigma$ comes from (\ref{eq:scalpot0})
\beq
m_\sigma^2= \frac{3}{2}\, \lambda \,f'^2\,.
\label{eq:msig}
\eeq
While there are also contributions to $m_\sigma^2$ proportional to
$f^2 \grotz^2$, it will be clear later that these are small corrections
compared to (\ref{eq:msig}) over the range of the parameters that will be
relevant to us. For the triplet pions, $m_\pi^2$ is generated
at ${\cal O}(\Grotz)$ from the piece of (\ref{eq:c1term}) that is
quadratic in $\pi_p$,
\beq
m_\pi^2=2c_1\sqrt{2}\frac{4\pi f}{f'} v^2 \grotz\,.
\label{eq:mpi}
\eeq
We will study (\ref{eq:msig}) and (\ref{eq:mpi}) quantitatively
in Section~\ref{sec:phenom}.

The reader should keep in mind that these results are not valid
everywhere in the model's parameter space. Since we want perturbation theory
to be reliable, we will always restrict ourselves to values of $\grotz$
and $\lambda$ that are less than $\sim 4\pi$ and $\sim 16\pi^2$,
respectively. However, chiral perturbation theory does not give an
appropriate description of the model everywhere in this
region. As we increase $\grotz$, the technifermion masses induced by
the scalar vev will eventually exceed the technicolor scale, and
chiral $SU(2)\times SU(2)$ will cease to be an approximate symmetry of the
theory. Since the technifermion masses are of order $\grotz f'$, the
effective chiral lagrangian is appropriate only if
\beq
\grotz f' \ll 4 \pi f\,,
\label{eq:inequal}
\eeq
or alternately,
\beq
\grotz \ll \left[\frac{2\sqrt{2}\pi^2}{c_1}\right]^{\frac{1}{4}}
\lambda^{\frac{1}{4}}\,.
\label{eq:bound}
\eeq

The physics of the model for $\grotz f' \gg 4 \pi f$ would be very peculiar.
The ``current'' masses of the technifermions produced by the breaking of the
$SU(2)\times U(1)$ would be much larger than the scale of the technicolor
interactions. From an effective field theory standpoint, you might worry that
the massive technifermions would not be present in the low energy theory to
produce $SU(2)\times U(1)$ breaking. But if there were no $SU(2)\times U(1)$
breaking, then the technifermions would be
massless and then they {\bf would} be in the low energy theory to produce
$SU(2)\times U(1)$ breaking! As we will see in the next section, the
Coleman-Weinberg
interactions save us from this logical conundrum. They push us away from
the region $\grotz f' \gg 4 \pi f$ toward the region (\ref{eq:bound}), where
the chiral Lagrangian description is valid. The corrections may be
important on the boundary of the allowed region. We will discuss the
phenomenology of this boundary region in Section~\ref{sec:beyond}

Note that all of the results presented in this section could have been
obtained had we chosen a different parametrization of the fields $\Sigma$
and $\Phi$. For example, we could have defined
\beq
\Phi = \frac{1}{\sqrt{2}}\Sigma_0 (f'+\sigma +2 i \alpha \pi_p)
\label{eq:newphidef}
\eeq
and
\beq
\Sigma = \Sigma_0 \exp(-2 i \beta \pi_p)
\label{eq:newsigdef}
\eeq
where
\begin{eqnarray}
\alpha=\frac{f}{v}&\;\mbox{,}\;&
\beta=\frac{f'}{f v}
\end{eqnarray}
and where
\beq
\Sigma_0 = \exp(\frac{2 i \pi_a}{v})
\label{eq:sig0def}
\eeq
with $f^2+f'^2=v^2$ as before. Unlike the field
redefinition given by (\ref{eq:redef1}), this choice is clearly
well-behaved as $f' \rightarrow 0$. While
(\ref{eq:redef1}) appears ill-defined in this limit, this
is simply the consequence of a coordinate singularity. The reader
can verify that either redefinition will yield the same physical results.

\section{Coleman-Weinberg Terms}\label{sec:cw}

In this section, we consider the effects of radiative corrections
to the $\sigma$ potential. The contributions will be important only when the
couplings involved can grow large somewhere in our parameter space. The only
Yukawa couplings that can become large are $\grotz_t$, and $\grotz_\pm$. Thus
to a good approximation, the corrected potential
can be written
\beq
V(\sigma)= \frac{\lambda}{8}\sigma^4
-\frac{1}{64\pi^2}\left(
3\grotz_t^4
+N(\grotz_+^4
+\grotz_-^4)\right)
\sigma^4
 \log \left(\frac{\sigma^2}{\mu^2}\right)
-8\sqrt{2}\pi c_1 f^3 \grotz
\sigma
\label{eq:cwpot}
\eeq
where $\grotz_t$ is the top quark Yukawa coupling, and $\mu$
is an arbitrary renormalization scale. To remove the $\mu$
dependence in (\ref{eq:cwpot}), we next define the renormalized
coupling $\lambda$ conventionally as,
\beq
\lambda\equiv\frac{1}{3}V''''(f')\,.
\label{eq:kbdef1}
\eeq
However, it will be convenient to describe the physics not with
(\ref{eq:kbdef1}), but in terms of the
parameter $\tilde\lambda$, defined by
\beq
\tilde{\lambda}\equiv\lambda + \frac{11}{24\pi^2} \left(
3\grotz_t^4+N(\grotz_+^4+\grotz_-^4)\right)
\,.
\label{eq:kbdef}
\eeq
We introduce the additional shift so that the condition $V'(f')=0$
gives us
\beq
f'=f(16\sqrt{2}\pi c_1)^{\frac{1}{3}}\left(\frac{\grotz}
{\tilde{\lambda}}\right)^{\frac{1}{3}}\,.
\label{eq:kbdef2}
\eeq
Because of (\ref{eq:kbdef2}), we can take over much of the analysis of the
previous section by simply making the replacement
$\lambda\rightarrow\tilde\lambda$.
For example, the
triplet pion mass contours, and the condition $\grotz f' < 4 \pi f$ look the
same as before, provided we use the parameter $\tilde{\lambda}$ instead
of $\lambda$.

Of course we cannot absorb all the effects of the radiative corrections
into a redefinition of $\lambda$. The one thing that does change is
the value of $m_\sigma$, which is determined by $V''(f')$. Expressed
in terms of $\tilde{\lambda}$, we find that
\beq
m^2_\sigma = \left(\frac{3}{2}\tilde{\lambda}-
\frac{1}{8\pi^2} \left(
3\grotz_t^4+N(\grotz_+^4+\grotz_-^4)\right)
\right) {f'}^2\,.
\label{eq:cwms}
\eeq
{}From (\ref{eq:cwms}), it is easy to see how they theory avoids the conundrum
described in the previous section. The stability of the vacuum requires
$m_\sigma^2\geq0$. Thus (\ref{eq:cwms}) implies
\beq
\frac{3}{2}\tilde{\lambda}\geq
\frac{N}{4\pi^2}(1+6\delta^2+\delta^4) \grotz^4\,.
\label{eq:stability}
\eeq
Together with (\ref{eq:vanish1}), this implies
\beq
\frac{\grotz f'}{4\pi f}\leq\left(\frac{3c_1}{\sqrt{2}
N(1+6\delta^2+\delta^4)}\right)^{1/3}\,.
\label{eq:stability2}
\eeq
Thus we never get into the dangerous region in which the effective field
theory description leads us into a logical puzzle. We therefore expect that
the chiral Lagrangian analysis of Section~\ref{sec:chiral} will give results
that are at least qualitatively correct. However, the right hand side of
(\ref{eq:stability2}) can be close to 1, so we may expect nontrivial
corrections to the chiral Lagrangian relations as we approach the line $\grotz
f'=4\pi f$.

\section{Flavor Changing Neutral Currents}\label{sec:fcnc}

In this section, we set up the analysis of flavor changing
neutral currents (FCNC) in our model. The coupling of the
scalar doublet $\phi$ to the ordinary quarks can be written
in matrix form
\beq
\overline{\psi}_L \Phi \left(\begin{array}{cc} \grotz_U & 0 \\
0 & V\grotz_D \end{array}\right) \psi_R + \mbox{h.c.}
\label{eq:qcoup}
\eeq
where $\psi_L = (U_L\mbox{,}V D_L)$, $\psi_R = (U_R\mbox{,}D_R)$,
$\grotz_U=$ diag($\grotz_u$,$\grotz_c$,$\grotz_t$),
$\grotz_D=$ diag($\grotz_d$,$\grotz_s$, $\grotz_b$),
and where $V$ is the CKM matrix. After applying (\ref{eq:redef1}),
these couplings become
\beq
\frac{(\sigma+f')}{\sqrt{2}} \overline{\psi}_L \Sigma '
\left(\begin{array}{cc} \grotz_U & 0 \\
0 & V\grotz_D \end{array}\right)\psi_R + \mbox{h.c.}
\eeq
from which we extract the charged-pion couplings
\beq
i(\frac{f}{v})\left[\overline{D}_L V^\dagger \pi_p^-\grotz_U U_R
+\overline{U}_L \pi^+_p V \grotz_D D_R + \mbox{h.c.}\right]\,.
\label{eq:onepicoup}
\eeq
\begin{figure}[htb]
\finput{$$
\beginpicture
\setcoordinatesystem units <\tdim,\tdim>
\put{\beginpicture
\setcoordinatesystem units <\tdim,\tdim>
\stpltsmbl
\linethickness=\tdim
\putrule from -90 30 to 90 30
\putrule from -90 -30 to 90 -30
\tarrow from -62 30 to -58 30
\tarrow from -2 30 to 2 30
\tarrow from 58 30 to 62 30
\tarrow from 62 -30 to 58 -30
\tarrow from 2 -30 to -2 -30
\tarrow from -58 -30 to -62 -30
\multiput {\copy\phdr}  at -30 30 *5 0 -10 /
\setdashes
\plot 30 30 30 -30 /
\put {$d$} at -100 30
\put {$q$} at 100 30
\put {$d$} at 100 -30
\put {$q$} at -100 -30
\put {$t$} at 0 -40
\put {$t$} at 0 40
\put {$W$} at -45 0
\put {$\pi_p$} at 43 0
\endpicture} at -120 0
\put {+} at 0 0
\put {\beginpicture
\setcoordinatesystem units <\tdim,\tdim>
\stpltsmbl
\linethickness=\tdim
\putrule from -90 30 to 90 30
\putrule from -90 -30 to 90 -30
\tarrow from -62 30 to -58 30
\tarrow from -2 30 to 2 30
\tarrow from 58 30 to 62 30
\tarrow from 62 -30 to 58 -30
\tarrow from 2 -30 to -2 -30
\tarrow from -58 -30 to -62 -30
\setdashes
\plot 30 30 30 -30 /
\plot -30 30 -30 -30 /
\put {$d$} at -100 30
\put {$q$} at 100 30
\put {$d$} at 100 -30
\put {$q$} at -100 -30
\put {$t$} at 0 -40
\put {$t$} at 0 40
\put {$\pi_p$} at -43 0
\put {$\pi_p$} at 43 0
\endpicture} at 120 0
\put {$(q=s,b)$} at 0 -70
\linethickness=0pt
\putrule from 0 -75 to 0 50
\endpicture
$$
}
\caption{Non-standard box diagrams for one and two pion exchange.}
\label{fig:boxdiag}
\end{figure}
The physical pions contribute to the $\Delta q=2$ box diagrams
shown in figure~\ref{fig:boxdiag}, where $q$=S or B\footnote{We
omit a discussion of $D^0$-$\overline{D^0}$ mixing, because we
found this process to provide much weaker constraints on the
model}. Notice that the explicit factors of the top quark Yukawa coupling
in these diagrams cause the two-top quark exchange diagrams to dominate
over all others. For the two-pion exchange diagram, we find that
the contribution to the operator
$(\overline{q} \gamma^\mu \frac{1+\gamma_5}{2} d)^2$
is given by
\beq
2\left(\frac{f}{f' v}\right)^4 m_t^4\, (V_{td} V^*_{tq})^2
\,I_1(m_t\mbox{,}m_\pi)
\label{eq:twopibox}
\eeq
while the one-pion-W exchange diagram gives us
\beq
g^2\left(\frac{f}{f' v}\right)^2 m_t^4\, (V_{td} V^*_{tq})^2
\left[I_2(m_t\mbox{,}m_\pi)+\frac{1}{4 m_W^2} I_3(m_t\mbox{,}m_\pi)
\right]\,.
\label{eq:onepibox}
\eeq
Similar box diagrams are generated in two-Higgs doublet models,
and the integrals $I_j$ ($j=1\cdots3$) have been tabulated
previously~\cite{wise}. We provide them in an appendix.
In evaluating these diagrams, we have neglected the
four-momenta of all external particles; this is a good approximation
because the masses of the $s$ and $b$ quarks are small compared to
the masses of the particles running around the loop.
For comparison, we note that the standard model contribution to
the coefficient of $(\overline{q} \gamma^\mu \frac{1+\gamma_5}{2} d)^2$
is of the order
\beq
\frac{G^2_F}{4\pi^2} m_c^2 (V_{cd} V_{cs}^*)^2
\label{eq:smk}
\eeq
for $K^0$-$\overline{K^0}$ mixing, and
\beq
\frac{G^2_F}{4\pi^2} m_t^2 (V_{td} V_{tb}^*)^2
\label{eq:smb}
\eeq
for $B^0$-$\overline{B^0}$ mixing. We will analyze these results
quantitatively in the following section.

\section{Experimental Limits}\label{sec:phenom}

We are now prepared to compare the predictions of our model
to experiment. As discussed in Section~\ref{sec:model},
we will strive to systematically exclude regions of
the $\lambda$-$\grotz$ plane. We immediately truncate our plots
at $\grotz=4\pi$ and $\lambda=16\pi^2$, and we show the
$\grotz f'= 4 \pi f$ line to indicate where there may be sizable corrections
to the chiral Lagrangian analysis.

Our first concern is that the scalar states in our model must be heavy
enough to avoid detection. The ALEPH Collaboration~\cite{aleph}
has placed the strongest lower limit on the mass of a neutral higgs
by studying the process $Z\rightarrow Z^* H$, assuming the standard model
coupling
\beq
\frac{v e^2}{4 s^2 c^2} Z^\mu Z_\mu H\,.
\eeq
They exclude a mass less than 48 GeV at the 95\% confidence
level. We display the $m_\sigma=48$ GeV contour in fig.~\ref{fig:plot0}. The
allowed region in inside the solid curve.
As discussed in the previous section, because of the Coleman-Weinberg terms in
the potential, $m_\sigma$ depends on $m_t$ and $\delta$. The solid line in
fig.~\ref{fig:plot0} corresponds to $m_t=110$~GeV and $\delta=0$. The Yukawa
coupling, $h_t$, gets large as $h$ gets small (see (\ref{eq:fprime})), thus
the effect of $m_t$ is large near the bottom of the $m_\sigma$ curve. Larger
values of $m_t$ move the bottom of the curve upwards. The effect of the
technifermion Yukawa couplings is large for large $h$, and thus the $\delta$
dependence is largest at the top of the curve. Increasing $\delta$ moves the
top of the curve down. The dashed line corresponds to $m_t=150$~GeV and
$\delta=0$.
\begin{figure}[htb]
\finput{$$
\beginpicture
\setcoordinatesystem units <90\tdim,60\tdim>
\setplotarea x from -2 to 2.2, y from -5 to 1.1
\linethickness=0pt
\putrule from 0 -5.3 to 0 1.7
\inboundscheckon
\linethickness=.013truein
\axis bottom ticks long in logged at .01 .1 1 10 100 / /
\axis top ticks long in logged at .01 .1 1 10 100 / /
\put {$10^{-2}$} [b] at -2 -5.35
\put {$10^{-1}$} [b] at -1 -5.35
\put {$1$} [b] at -0 -5.35
\put {$2$} [b] at .3 -5.35
\put {$5$} [b] at .7 -5.35
\put {$10$} [b] at 1 -5.35
\put {$10^2$} [b] at 2 -5.35
\linethickness=.006truein
\axis bottom ticks short in logged unlabeled at .02 .2 2 20 / /
\axis bottom ticks short in logged unlabeled at .03 .3 3 30 / /
\axis bottom ticks short in logged unlabeled at .04 .4 4 40 / /
\axis bottom ticks short in logged unlabeled at .05 .5 5 50 / /
\axis top ticks short in logged unlabeled at .02 .2 2 20 / /
\axis top ticks short in logged unlabeled at .03 .3 3 30 / /
\axis top ticks short in logged unlabeled at .04 .4 4 40 / /
\axis top ticks short in logged unlabeled at .05 .5 5 50 / /
\linethickness=.013truein
\put {$10^{-5}$} [r] at -2.08 -5
\put {$10^{-4}$} [r] at -2.08 -4
\put {$10^{-3}$} [r] at -2.08 -3
\put {$10^{-2}$} [r] at -2.08 -2
\put {$10^{-1}$} [r] at -2.08 -1
\put {$1$} [r] at -2.08 -0
\put {$2$} [r] at -2.08 .3
\put {$5$} [r] at -2.08 .7
\put {$10$} [r] at -2.08 1
\axis right ticks long in logged unlabeled at .00001 .0001 .001 .01 .1 1  10 /
/
\axis left ticks long in logged unlabeled at .00001 .0001 .001 .01 .1 1 10 / /
\linethickness=.006truein
\axis left ticks short in logged unlabeled at .00002 .0002 .002 .02 .2 2 / /
\axis left ticks short in logged unlabeled at .00003 .0003 .003 .03 .3 3 / /
\axis left ticks short in logged unlabeled at .00004 .0004 .004 .04 .4 4 / /
\axis left ticks short in logged unlabeled at .00005 .0005 .005 .05 .5 5 / /
\axis right ticks short in logged unlabeled at .00002 .0002 .002 .02 .2 2 / /
\axis right ticks short in logged unlabeled at .00003 .0003 .003 .03 .3 3 / /
\axis right ticks short in logged unlabeled at .00004 .0004 .004 .04 .4 4 / /
\axis right ticks short in logged unlabeled at .00005 .0005 .005 .05 .5 5 / /
\put {$\tilde\lambda\rightarrow$} at .1 -5.6
\put {\stack{$\uparrow$,$h$}} at -2.5 -2
\plot
  3.017  -3.356
  2.007  -3.589
  1.554  -3.683
  1.259  -3.738
  1.039  -3.774
  0.864  -3.798
  0.719  -3.815
  0.594  -3.827
  0.486  -3.835
  0.389  -3.840
  0.302  -3.843
  0.223  -3.844
  0.151  -3.843
  0.084  -3.840
  0.022  -3.837
 -0.036  -3.832
 -0.091  -3.826
 -0.142  -3.820
 -0.191  -3.813
 -0.237  -3.806
 -0.281  -3.797
 -0.322  -3.789
 -0.362  -3.779
 -0.400  -3.770
 -0.437  -3.760
 -0.472  -3.749
 -0.506  -3.739
 -0.538  -3.728
 -0.570  -3.716
 -0.600  -3.705
 -0.629  -3.693
 -0.657  -3.681
 -0.684  -3.669
 -0.711  -3.656
 -0.736  -3.643
 -0.761  -3.630
 -0.785  -3.617
 -0.808  -3.604
 -0.831  -3.590
 -0.853  -3.577
 -0.874  -3.563
 -0.895  -3.549
 -0.915  -3.535
 -0.934  -3.521
 -0.953  -3.506
 -0.972  -3.492
 -0.990  -3.477
 -1.007  -3.462
 -1.024  -3.447
 -1.040  -3.432
 -1.056  -3.417
 -1.071  -3.402
 -1.086  -3.386
 -1.101  -3.371
 -1.115  -3.355
 -1.129  -3.339
 -1.142  -3.324
 -1.155  -3.308
 -1.168  -3.292
 -1.180  -3.276
 -1.192  -3.260
 -1.204  -3.243
 -1.215  -3.227
 -1.226  -3.211
 -1.237  -3.194
 -1.247  -3.178
 -1.257  -3.161
 -1.267  -3.144
 -1.276  -3.127
 -1.286  -3.111
 -1.295  -3.094
 -1.303  -3.077
 -1.312  -3.060
 -1.320  -3.043
 -1.328  -3.025
 -1.335  -3.008
 -1.343  -2.991
 -1.350  -2.974
 -1.357  -2.956
 -1.364  -2.939
 -1.371  -2.921
 -1.377  -2.904
 -1.384  -2.886
 -1.390  -2.869
 -1.396  -2.851
 -1.401  -2.833
 -1.407  -2.816
 -1.412  -2.798
 -1.418  -2.780
 -1.423  -2.762
 -1.428  -2.744
 -1.433  -2.726
 -1.437  -2.708
 -1.442  -2.690
 -1.446  -2.672
 -1.450  -2.654
 -1.455  -2.636
 -1.459  -2.618
 -1.463  -2.600
 -1.466  -2.582
 -1.470  -2.563
 -1.474  -2.545
 -1.477  -2.527
 -1.480  -2.509
 -1.484  -2.490
 -1.487  -2.472
 -1.490  -2.453
 -1.493  -2.435
 -1.496  -2.417
 -1.499  -2.398
 -1.501  -2.380
 -1.504  -2.361
 -1.507  -2.343
 -1.509  -2.324
 -1.512  -2.306
 -1.514  -2.287
 -1.516  -2.269
 -1.519  -2.250
 -1.521  -2.231
 -1.523  -2.213
 -1.525  -2.194
 -1.527  -2.176
 -1.529  -2.157
 -1.531  -2.138
 -1.532  -2.120
 -1.534  -2.101
 -1.536  -2.082
 -1.538  -2.063
 -1.539  -2.045
 -1.541  -2.026
 -1.542  -2.007
 -1.544  -1.988
 -1.545  -1.970
 -1.547  -1.951
 -1.548  -1.932
 -1.549  -1.913
 -1.551  -1.894
 -1.552  -1.876
 -1.553  -1.857
 -1.554  -1.838
 -1.555  -1.819
 -1.556  -1.800
 -1.557  -1.781
 -1.558  -1.762
 -1.559  -1.743
 -1.560  -1.725
 -1.561  -1.706
 -1.562  -1.687
 -1.563  -1.668
 -1.564  -1.649
 -1.565  -1.630
 -1.566  -1.611
 -1.566  -1.592
 -1.567  -1.573
 -1.568  -1.554
 -1.569  -1.535
 -1.569  -1.516
 -1.570  -1.497
 -1.571  -1.478
 -1.571  -1.459
 -1.572  -1.440
 -1.573  -1.421
 -1.573  -1.402
 -1.574  -1.383
 -1.574  -1.364
 -1.575  -1.345
 -1.575  -1.326
 -1.576  -1.307
 -1.576  -1.288
 -1.577  -1.269
 -1.577  -1.250
 -1.578  -1.231
 -1.578  -1.212
 -1.579  -1.193
 -1.579  -1.174
 -1.580  -1.155
 -1.580  -1.136
 -1.580  -1.117
 -1.581  -1.098
 -1.581  -1.079
 -1.581  -1.060
 -1.582  -1.041
 -1.582  -1.022
 -1.582  -1.003
 -1.583  -0.984
 -1.583  -0.965
 -1.583  -0.945
 -1.583  -0.926
 -1.584  -0.907
 -1.584  -0.888
 -1.584  -0.869
 -1.584  -0.850
 -1.584  -0.831
 -1.584  -0.812
 -1.585  -0.793
 -1.585  -0.774
 -1.585  -0.754
 -1.585  -0.735
 -1.585  -0.716
 -1.585  -0.697
 -1.584  -0.678
 -1.584  -0.659
 -1.584  -0.639
 -1.583  -0.620
 -1.583  -0.601
 -1.582  -0.581
 -1.581  -0.562
 -1.580  -0.543
 -1.579  -0.523
 -1.577  -0.504
 -1.575  -0.484
 -1.573  -0.464
 -1.570  -0.444
 -1.566  -0.424
 -1.562  -0.404
 -1.557  -0.384
 -1.550  -0.363
 -1.543  -0.342
 -1.533  -0.320
 -1.522  -0.298
 -1.507  -0.276
 -1.490  -0.252
 -1.468  -0.227
 -1.440  -0.201
 -1.404  -0.173
 -1.357  -0.142
 -1.293  -0.107
 -1.202  -0.065
 -1.061  -0.011
 -0.795   0.075
  2.449   0.905
/
\setdashes <5pt>
\plot
  2.653  -3.043
  1.714  -3.261
  1.277  -3.354
  0.990  -3.410
  0.776  -3.447
  0.606  -3.473
  0.465  -3.491
  0.345  -3.505
  0.241  -3.515
  0.148  -3.522
  0.065  -3.526
 -0.010  -3.528
 -0.078  -3.529
 -0.141  -3.528
 -0.199  -3.526
 -0.253  -3.523
 -0.303  -3.520
 -0.350  -3.515
 -0.395  -3.510
 -0.436  -3.504
 -0.476  -3.497
 -0.513  -3.490
 -0.548  -3.482
 -0.582  -3.474
 -0.614  -3.466
 -0.644  -3.457
 -0.674  -3.448
 -0.702  -3.439
 -0.728  -3.429
 -0.754  -3.419
 -0.778  -3.409
 -0.802  -3.398
 -0.825  -3.387
 -0.846  -3.377
 -0.867  -3.365
 -0.888  -3.354
 -0.907  -3.342
 -0.926  -3.331
 -0.944  -3.319
 -0.962  -3.307
 -0.979  -3.295
 -0.995  -3.282
 -1.011  -3.270
 -1.026  -3.257
 -1.041  -3.245
 -1.055  -3.232
 -1.069  -3.219
 -1.083  -3.206
 -1.096  -3.193
 -1.108  -3.180
 -1.121  -3.166
 -1.133  -3.153
 -1.144  -3.139
 -1.155  -3.126
 -1.166  -3.112
 -1.177  -3.098
 -1.187  -3.084
 -1.197  -3.070
 -1.206  -3.056
 -1.216  -3.042
 -1.225  -3.028
 -1.234  -3.014
 -1.242  -3.000
 -1.250  -2.985
 -1.258  -2.971
 -1.266  -2.957
 -1.274  -2.942
 -1.281  -2.927
 -1.289  -2.913
 -1.296  -2.898
 -1.302  -2.884
 -1.309  -2.869
 -1.315  -2.854
 -1.322  -2.839
 -1.328  -2.824
 -1.334  -2.809
 -1.339  -2.794
 -1.345  -2.779
 -1.351  -2.764
 -1.356  -2.749
 -1.361  -2.734
 -1.366  -2.719
 -1.371  -2.704
 -1.376  -2.689
 -1.380  -2.673
 -1.385  -2.658
 -1.389  -2.643
 -1.393  -2.627
 -1.398  -2.612
 -1.402  -2.597
 -1.406  -2.581
 -1.409  -2.566
 -1.413  -2.550
 -1.417  -2.535
 -1.420  -2.519
 -1.424  -2.504
 -1.427  -2.488
 -1.430  -2.473
 -1.434  -2.457
 -1.437  -2.441
 -1.440  -2.426
 -1.443  -2.410
 -1.445  -2.394
 -1.448  -2.379
 -1.451  -2.363
 -1.454  -2.347
 -1.456  -2.331
 -1.459  -2.316
 -1.461  -2.300
 -1.464  -2.284
 -1.466  -2.268
 -1.468  -2.252
 -1.470  -2.237
 -1.472  -2.221
 -1.475  -2.205
 -1.477  -2.189
 -1.479  -2.173
 -1.481  -2.157
 -1.482  -2.141
 -1.484  -2.125
 -1.486  -2.109
 -1.488  -2.093
 -1.489  -2.077
 -1.491  -2.061
 -1.493  -2.045
 -1.494  -2.029
 -1.496  -2.013
 -1.497  -1.997
 -1.499  -1.981
 -1.500  -1.965
 -1.502  -1.949
 -1.503  -1.933
 -1.504  -1.917
 -1.505  -1.901
 -1.507  -1.885
 -1.508  -1.869
 -1.509  -1.852
 -1.510  -1.836
 -1.511  -1.820
 -1.512  -1.804
 -1.514  -1.788
 -1.515  -1.772
 -1.516  -1.756
 -1.517  -1.740
 -1.518  -1.723
 -1.519  -1.707
 -1.519  -1.691
 -1.520  -1.675
 -1.521  -1.659
 -1.522  -1.642
 -1.523  -1.626
 -1.524  -1.610
 -1.524  -1.594
 -1.525  -1.578
 -1.526  -1.561
 -1.527  -1.545
 -1.527  -1.529
 -1.528  -1.513
 -1.529  -1.497
 -1.529  -1.480
 -1.530  -1.464
 -1.531  -1.448
 -1.531  -1.432
 -1.532  -1.415
 -1.532  -1.399
 -1.533  -1.383
 -1.534  -1.366
 -1.534  -1.350
 -1.535  -1.334
 -1.535  -1.318
 -1.536  -1.301
 -1.536  -1.285
 -1.537  -1.269
 -1.537  -1.252
 -1.537  -1.236
 -1.538  -1.220
 -1.538  -1.204
 -1.539  -1.187
 -1.539  -1.171
 -1.539  -1.155
 -1.540  -1.138
 -1.540  -1.122
 -1.540  -1.106
 -1.541  -1.089
 -1.541  -1.073
 -1.541  -1.057
 -1.542  -1.040
 -1.542  -1.024
 -1.542  -1.008
 -1.542  -0.991
 -1.542  -0.975
 -1.542  -0.959
 -1.543  -0.942
 -1.543  -0.926
 -1.543  -0.909
 -1.543  -0.893
 -1.542  -0.877
 -1.542  -0.860
 -1.542  -0.844
 -1.542  -0.827
 -1.541  -0.811
 -1.541  -0.794
 -1.540  -0.778
 -1.539  -0.761
 -1.538  -0.744
 -1.537  -0.728
 -1.535  -0.711
 -1.534  -0.694
 -1.532  -0.677
 -1.529  -0.660
 -1.526  -0.643
 -1.523  -0.626
 -1.518  -0.608
 -1.514  -0.590
 -1.508  -0.573
 -1.501  -0.554
 -1.493  -0.536
 -1.483  -0.517
 -1.471  -0.498
 -1.457  -0.478
 -1.441  -0.457
 -1.420  -0.436
 -1.395  -0.413
 -1.364  -0.389
 -1.324  -0.363
 -1.273  -0.333
 -1.205  -0.300
 -1.109  -0.260
 -0.963  -0.207
 -0.691  -0.122
  2.691   0.740
/
\setdots <2pt>
\plot
 -2.000  -0.139
  0.100   0.386
  2.200   0.911
/
\endpicture
$$
}
\caption{The $m_\sigma=48$ GeV line. Solid line is $m_t=110$~GeV, $\delta=0$.
Dashed line is $m_t=150$~GeV, $\delta=1$. The dotted line is $\grotz f'=4\pi
f$.}
\label{fig:plot0}
\end{figure}
It is important to keep in mind that the $ZZ\sigma$
coupling in our model is given by
\beq
\frac{f' e^2}{4 s^2 c^2} Z^\mu Z_\mu \sigma
\eeq
and hence the $\sigma$ production rate is suppressed compared
to the standard model by a factor of $(f'/v)^2$. Thus, the 48 GeV
contour is not equivalent to the LEP limit everywhere. However,
we will see shortly out that in the region of parameter space not
excluded by flavor changing neutral currents, $f'/v\sim 1$ along
the $m_\sigma=$48 GeV line, and it therefore serves as an approximate
boundary to the experimentally allowed region.

\begin{figure}[htb]
\finput{$$
\beginpicture
\setcoordinatesystem units <90\tdim,60\tdim>
\setplotarea x from -2 to 2.2, y from -5 to 1.1
\linethickness=0pt
\putrule from 0 -5.3 to 0 1.7
\inboundscheckon
\linethickness=.013truein
\axis bottom ticks long in logged at .01 .1 1 10 100 / /
\axis top ticks long in logged at .01 .1 1 10 100 / /
\put {$10^{-2}$} [b] at -2 -5.35
\put {$10^{-1}$} [b] at -1 -5.35
\put {$1$} [b] at -0 -5.35
\put {$2$} [b] at .3 -5.35
\put {$5$} [b] at .7 -5.35
\put {$10$} [b] at 1 -5.35
\put {$10^2$} [b] at 2 -5.35
\linethickness=.006truein
\axis bottom ticks short in logged unlabeled at .02 .2 2 20 / /
\axis bottom ticks short in logged unlabeled at .03 .3 3 30 / /
\axis bottom ticks short in logged unlabeled at .04 .4 4 40 / /
\axis bottom ticks short in logged unlabeled at .05 .5 5 50 / /
\axis top ticks short in logged unlabeled at .02 .2 2 20 / /
\axis top ticks short in logged unlabeled at .03 .3 3 30 / /
\axis top ticks short in logged unlabeled at .04 .4 4 40 / /
\axis top ticks short in logged unlabeled at .05 .5 5 50 / /
\linethickness=.013truein
\put {$10^{-5}$} [r] at -2.08 -5
\put {$10^{-4}$} [r] at -2.08 -4
\put {$10^{-3}$} [r] at -2.08 -3
\put {$10^{-2}$} [r] at -2.08 -2
\put {$10^{-1}$} [r] at -2.08 -1
\put {$1$} [r] at -2.08 -0
\put {$2$} [r] at -2.08 .3
\put {$5$} [r] at -2.08 .7
\put {$10$} [r] at -2.08 1
\axis right ticks long in logged unlabeled at .00001 .0001 .001 .01 .1 1  10 /
/
\axis left ticks long in logged unlabeled at .00001 .0001 .001 .01 .1 1 10 / /
\linethickness=.006truein
\axis left ticks short in logged unlabeled at .00002 .0002 .002 .02 .2 2 / /
\axis left ticks short in logged unlabeled at .00003 .0003 .003 .03 .3 3 / /
\axis left ticks short in logged unlabeled at .00004 .0004 .004 .04 .4 4 / /
\axis left ticks short in logged unlabeled at .00005 .0005 .005 .05 .5 5 / /
\axis right ticks short in logged unlabeled at .00002 .0002 .002 .02 .2 2 / /
\axis right ticks short in logged unlabeled at .00003 .0003 .003 .03 .3 3 / /
\axis right ticks short in logged unlabeled at .00004 .0004 .004 .04 .4 4 / /
\axis right ticks short in logged unlabeled at .00005 .0005 .005 .05 .5 5 / /
\put {$\tilde\lambda\rightarrow$} at .1 -5.6
\put {\stack{$\uparrow$,$h$}} at -2.5 -2
\plot
  3.017  -3.356
  2.007  -3.589
  1.554  -3.683
  1.259  -3.738
  1.039  -3.774
  0.864  -3.798
  0.719  -3.815
  0.594  -3.827
  0.486  -3.835
  0.389  -3.840
  0.302  -3.843
  0.223  -3.844
  0.151  -3.843
  0.084  -3.840
  0.022  -3.837
 -0.036  -3.832
 -0.091  -3.826
 -0.142  -3.820
 -0.191  -3.813
 -0.237  -3.806
 -0.281  -3.797
 -0.322  -3.789
 -0.362  -3.779
 -0.400  -3.770
 -0.437  -3.760
 -0.472  -3.749
 -0.506  -3.739
 -0.538  -3.728
 -0.570  -3.716
 -0.600  -3.705
 -0.629  -3.693
 -0.657  -3.681
 -0.684  -3.669
 -0.711  -3.656
 -0.736  -3.643
 -0.761  -3.630
 -0.785  -3.617
 -0.808  -3.604
 -0.831  -3.590
 -0.853  -3.577
 -0.874  -3.563
 -0.895  -3.549
 -0.915  -3.535
 -0.934  -3.521
 -0.953  -3.506
 -0.972  -3.492
 -0.990  -3.477
 -1.007  -3.462
 -1.024  -3.447
 -1.040  -3.432
 -1.056  -3.417
 -1.071  -3.402
 -1.086  -3.386
 -1.101  -3.371
 -1.115  -3.355
 -1.129  -3.339
 -1.142  -3.324
 -1.155  -3.308
 -1.168  -3.292
 -1.180  -3.276
 -1.192  -3.260
 -1.204  -3.243
 -1.215  -3.227
 -1.226  -3.211
 -1.237  -3.194
 -1.247  -3.178
 -1.257  -3.161
 -1.267  -3.144
 -1.276  -3.127
 -1.286  -3.111
 -1.295  -3.094
 -1.303  -3.077
 -1.312  -3.060
 -1.320  -3.043
 -1.328  -3.025
 -1.335  -3.008
 -1.343  -2.991
 -1.350  -2.974
 -1.357  -2.956
 -1.364  -2.939
 -1.371  -2.921
 -1.377  -2.904
 -1.384  -2.886
 -1.390  -2.869
 -1.396  -2.851
 -1.401  -2.833
 -1.407  -2.816
 -1.412  -2.798
 -1.418  -2.780
 -1.423  -2.762
 -1.428  -2.744
 -1.433  -2.726
 -1.437  -2.708
 -1.442  -2.690
 -1.446  -2.672
 -1.450  -2.654
 -1.455  -2.636
 -1.459  -2.618
 -1.463  -2.600
 -1.466  -2.582
 -1.470  -2.563
 -1.474  -2.545
 -1.477  -2.527
 -1.480  -2.509
 -1.484  -2.490
 -1.487  -2.472
 -1.490  -2.453
 -1.493  -2.435
 -1.496  -2.417
 -1.499  -2.398
 -1.501  -2.380
 -1.504  -2.361
 -1.507  -2.343
 -1.509  -2.324
 -1.512  -2.306
 -1.514  -2.287
 -1.516  -2.269
 -1.519  -2.250
 -1.521  -2.231
 -1.523  -2.213
 -1.525  -2.194
 -1.527  -2.176
 -1.529  -2.157
 -1.531  -2.138
 -1.532  -2.120
 -1.534  -2.101
 -1.536  -2.082
 -1.538  -2.063
 -1.539  -2.045
 -1.541  -2.026
 -1.542  -2.007
 -1.544  -1.988
 -1.545  -1.970
 -1.547  -1.951
 -1.548  -1.932
 -1.549  -1.913
 -1.551  -1.894
 -1.552  -1.876
 -1.553  -1.857
 -1.554  -1.838
 -1.555  -1.819
 -1.556  -1.800
 -1.557  -1.781
 -1.558  -1.762
 -1.559  -1.743
 -1.560  -1.725
 -1.561  -1.706
 -1.562  -1.687
 -1.563  -1.668
 -1.564  -1.649
 -1.565  -1.630
 -1.566  -1.611
 -1.566  -1.592
 -1.567  -1.573
 -1.568  -1.554
 -1.569  -1.535
 -1.569  -1.516
 -1.570  -1.497
 -1.571  -1.478
 -1.571  -1.459
 -1.572  -1.440
 -1.573  -1.421
 -1.573  -1.402
 -1.574  -1.383
 -1.574  -1.364
 -1.575  -1.345
 -1.575  -1.326
 -1.576  -1.307
 -1.576  -1.288
 -1.577  -1.269
 -1.577  -1.250
 -1.578  -1.231
 -1.578  -1.212
 -1.579  -1.193
 -1.579  -1.174
 -1.580  -1.155
 -1.580  -1.136
 -1.580  -1.117
 -1.581  -1.098
 -1.581  -1.079
 -1.581  -1.060
 -1.582  -1.041
 -1.582  -1.022
 -1.582  -1.003
 -1.583  -0.984
 -1.583  -0.965
 -1.583  -0.945
 -1.583  -0.926
 -1.584  -0.907
 -1.584  -0.888
 -1.584  -0.869
 -1.584  -0.850
 -1.584  -0.831
 -1.584  -0.812
 -1.585  -0.793
 -1.585  -0.774
 -1.585  -0.754
 -1.585  -0.735
 -1.585  -0.716
 -1.585  -0.697
 -1.584  -0.678
 -1.584  -0.659
 -1.584  -0.639
 -1.583  -0.620
 -1.583  -0.601
 -1.582  -0.581
 -1.581  -0.562
 -1.580  -0.543
 -1.579  -0.523
 -1.577  -0.504
 -1.575  -0.484
 -1.573  -0.464
 -1.570  -0.444
 -1.566  -0.424
 -1.562  -0.404
 -1.557  -0.384
 -1.550  -0.363
 -1.543  -0.342
 -1.533  -0.320
 -1.522  -0.298
 -1.507  -0.276
 -1.490  -0.252
 -1.468  -0.227
 -1.440  -0.201
 -1.404  -0.173
 -1.357  -0.142
 -1.293  -0.107
 -1.202  -0.065
 -1.061  -0.011
 -0.795   0.075
  2.449   0.905
/
\put {$m_\sigma=48$~GeV} [l] at -1.4 -1.5
\setdots <2pt>
\plot
 -2.000  -0.139
  0.100   0.386
  2.200   0.911
/
\setsolid
\put {$m_t=110$~GeV} at 1 -3.25
\plot
 -2.000  -5.130
 -1.790  -4.924
 -1.580  -4.721
 -1.370  -4.520
 -1.160  -4.322
 -0.950  -4.128
 -0.740  -3.938
 -0.530  -3.754
 -0.320  -3.575
 -0.110  -3.403
  0.100  -3.237
  0.310  -3.077
  0.520  -2.924
  0.730  -2.777
  0.940  -2.636
  1.150  -2.500
  1.360  -2.369
  1.570  -2.241
  1.780  -2.118
  1.990  -1.997
  2.200  -1.880
/
\setdashes
\put {$m_t=150$~GeV} at 1 -1.75
\plot
 -2.000  -4.917
 -1.790  -4.711
 -1.580  -4.506
 -1.370  -4.304
 -1.160  -4.104
 -0.950  -3.908
 -0.740  -3.715
 -0.530  -3.528
 -0.320  -3.346
 -0.110  -3.170
  0.100  -3.000
  0.310  -2.837
  0.520  -2.680
  0.730  -2.530
  0.940  -2.385
  1.150  -2.246
  1.360  -2.112
  1.570  -1.982
  1.780  -1.856
  1.990  -1.734
  2.200  -1.615
/
\endpicture
$$
}
\caption{Parameter space excluded by $K^0$-$\overline{K^0}$
mixing; the region above the K-line
and to the right of the 48 GeV contour is allowed.}
\label{fig:plot1}
\end{figure}
In figs.~\ref{fig:plot1}
and~\ref{fig:plot2}, we show a rough estimate of the flavor changing neutral
current limits, generated by requiring that the sum of the
nonstandard box diagrams, (\ref{eq:twopibox}) and (\ref{eq:onepibox}),
not exceed the standard model estimates given in (\ref{eq:smk})
and (\ref{eq:smb}). We show the results for two representative values of
the top quark mass, $m_t=110$ and $150$ GeV. Evidently, as the experimental
lower bound on $m_t$ increases, the flavor changing neutral current limits
tighten.
The region in parameter space excluded by $K^0$-$\overline{K^0}$ and
$B^0$-$\overline{B^0}$ mixing lies below the `K-line' and `B-line',
respectively. We see that $B^0$-$\overline{B^0}$ mixing provides the
more stringent limits, but by no means excludes all of the available
parameter space.  Everything above the B-line, and to the right
of the $m_\sigma=48$ GeV contour is allowed, and much of this
parameter space is within the region in which our chiral expansion
is trustworthy.

\begin{figure}[htb]
\finput{$$
\beginpicture
\setcoordinatesystem units <90\tdim,60\tdim>
\setplotarea x from -2 to 2.2, y from -5 to 1.1
\linethickness=0pt
\putrule from 0 -5.3 to 0 1.7
\inboundscheckon
\linethickness=.013truein
\axis bottom ticks long in logged at .01 .1 1 10 100 / /
\axis top ticks long in logged at .01 .1 1 10 100 / /
\put {$10^{-2}$} [b] at -2 -5.35
\put {$10^{-1}$} [b] at -1 -5.35
\put {$1$} [b] at -0 -5.35
\put {$2$} [b] at .3 -5.35
\put {$5$} [b] at .7 -5.35
\put {$10$} [b] at 1 -5.35
\put {$10^2$} [b] at 2 -5.35
\linethickness=.006truein
\axis bottom ticks short in logged unlabeled at .02 .2 2 20 / /
\axis bottom ticks short in logged unlabeled at .03 .3 3 30 / /
\axis bottom ticks short in logged unlabeled at .04 .4 4 40 / /
\axis bottom ticks short in logged unlabeled at .05 .5 5 50 / /
\axis top ticks short in logged unlabeled at .02 .2 2 20 / /
\axis top ticks short in logged unlabeled at .03 .3 3 30 / /
\axis top ticks short in logged unlabeled at .04 .4 4 40 / /
\axis top ticks short in logged unlabeled at .05 .5 5 50 / /
\linethickness=.013truein
\put {$10^{-5}$} [r] at -2.08 -5
\put {$10^{-4}$} [r] at -2.08 -4
\put {$10^{-3}$} [r] at -2.08 -3
\put {$10^{-2}$} [r] at -2.08 -2
\put {$10^{-1}$} [r] at -2.08 -1
\put {$1$} [r] at -2.08 -0
\put {$2$} [r] at -2.08 .3
\put {$5$} [r] at -2.08 .7
\put {$10$} [r] at -2.08 1
\axis right ticks long in logged unlabeled at .00001 .0001 .001 .01 .1 1  10 /
/
\axis left ticks long in logged unlabeled at .00001 .0001 .001 .01 .1 1 10 / /
\linethickness=.006truein
\axis left ticks short in logged unlabeled at .00002 .0002 .002 .02 .2 2 / /
\axis left ticks short in logged unlabeled at .00003 .0003 .003 .03 .3 3 / /
\axis left ticks short in logged unlabeled at .00004 .0004 .004 .04 .4 4 / /
\axis left ticks short in logged unlabeled at .00005 .0005 .005 .05 .5 5 / /
\axis right ticks short in logged unlabeled at .00002 .0002 .002 .02 .2 2 / /
\axis right ticks short in logged unlabeled at .00003 .0003 .003 .03 .3 3 / /
\axis right ticks short in logged unlabeled at .00004 .0004 .004 .04 .4 4 / /
\axis right ticks short in logged unlabeled at .00005 .0005 .005 .05 .5 5 / /
\put {$\tilde\lambda\rightarrow$} at .1 -5.6
\put {\stack{$\uparrow$,$h$}} at -2.5 -2
\plot
  3.017  -3.356
  2.007  -3.589
  1.554  -3.683
  1.259  -3.738
  1.039  -3.774
  0.864  -3.798
  0.719  -3.815
  0.594  -3.827
  0.486  -3.835
  0.389  -3.840
  0.302  -3.843
  0.223  -3.844
  0.151  -3.843
  0.084  -3.840
  0.022  -3.837
 -0.036  -3.832
 -0.091  -3.826
 -0.142  -3.820
 -0.191  -3.813
 -0.237  -3.806
 -0.281  -3.797
 -0.322  -3.789
 -0.362  -3.779
 -0.400  -3.770
 -0.437  -3.760
 -0.472  -3.749
 -0.506  -3.739
 -0.538  -3.728
 -0.570  -3.716
 -0.600  -3.705
 -0.629  -3.693
 -0.657  -3.681
 -0.684  -3.669
 -0.711  -3.656
 -0.736  -3.643
 -0.761  -3.630
 -0.785  -3.617
 -0.808  -3.604
 -0.831  -3.590
 -0.853  -3.577
 -0.874  -3.563
 -0.895  -3.549
 -0.915  -3.535
 -0.934  -3.521
 -0.953  -3.506
 -0.972  -3.492
 -0.990  -3.477
 -1.007  -3.462
 -1.024  -3.447
 -1.040  -3.432
 -1.056  -3.417
 -1.071  -3.402
 -1.086  -3.386
 -1.101  -3.371
 -1.115  -3.355
 -1.129  -3.339
 -1.142  -3.324
 -1.155  -3.308
 -1.168  -3.292
 -1.180  -3.276
 -1.192  -3.260
 -1.204  -3.243
 -1.215  -3.227
 -1.226  -3.211
 -1.237  -3.194
 -1.247  -3.178
 -1.257  -3.161
 -1.267  -3.144
 -1.276  -3.127
 -1.286  -3.111
 -1.295  -3.094
 -1.303  -3.077
 -1.312  -3.060
 -1.320  -3.043
 -1.328  -3.025
 -1.335  -3.008
 -1.343  -2.991
 -1.350  -2.974
 -1.357  -2.956
 -1.364  -2.939
 -1.371  -2.921
 -1.377  -2.904
 -1.384  -2.886
 -1.390  -2.869
 -1.396  -2.851
 -1.401  -2.833
 -1.407  -2.816
 -1.412  -2.798
 -1.418  -2.780
 -1.423  -2.762
 -1.428  -2.744
 -1.433  -2.726
 -1.437  -2.708
 -1.442  -2.690
 -1.446  -2.672
 -1.450  -2.654
 -1.455  -2.636
 -1.459  -2.618
 -1.463  -2.600
 -1.466  -2.582
 -1.470  -2.563
 -1.474  -2.545
 -1.477  -2.527
 -1.480  -2.509
 -1.484  -2.490
 -1.487  -2.472
 -1.490  -2.453
 -1.493  -2.435
 -1.496  -2.417
 -1.499  -2.398
 -1.501  -2.380
 -1.504  -2.361
 -1.507  -2.343
 -1.509  -2.324
 -1.512  -2.306
 -1.514  -2.287
 -1.516  -2.269
 -1.519  -2.250
 -1.521  -2.231
 -1.523  -2.213
 -1.525  -2.194
 -1.527  -2.176
 -1.529  -2.157
 -1.531  -2.138
 -1.532  -2.120
 -1.534  -2.101
 -1.536  -2.082
 -1.538  -2.063
 -1.539  -2.045
 -1.541  -2.026
 -1.542  -2.007
 -1.544  -1.988
 -1.545  -1.970
 -1.547  -1.951
 -1.548  -1.932
 -1.549  -1.913
 -1.551  -1.894
 -1.552  -1.876
 -1.553  -1.857
 -1.554  -1.838
 -1.555  -1.819
 -1.556  -1.800
 -1.557  -1.781
 -1.558  -1.762
 -1.559  -1.743
 -1.560  -1.725
 -1.561  -1.706
 -1.562  -1.687
 -1.563  -1.668
 -1.564  -1.649
 -1.565  -1.630
 -1.566  -1.611
 -1.566  -1.592
 -1.567  -1.573
 -1.568  -1.554
 -1.569  -1.535
 -1.569  -1.516
 -1.570  -1.497
 -1.571  -1.478
 -1.571  -1.459
 -1.572  -1.440
 -1.573  -1.421
 -1.573  -1.402
 -1.574  -1.383
 -1.574  -1.364
 -1.575  -1.345
 -1.575  -1.326
 -1.576  -1.307
 -1.576  -1.288
 -1.577  -1.269
 -1.577  -1.250
 -1.578  -1.231
 -1.578  -1.212
 -1.579  -1.193
 -1.579  -1.174
 -1.580  -1.155
 -1.580  -1.136
 -1.580  -1.117
 -1.581  -1.098
 -1.581  -1.079
 -1.581  -1.060
 -1.582  -1.041
 -1.582  -1.022
 -1.582  -1.003
 -1.583  -0.984
 -1.583  -0.965
 -1.583  -0.945
 -1.583  -0.926
 -1.584  -0.907
 -1.584  -0.888
 -1.584  -0.869
 -1.584  -0.850
 -1.584  -0.831
 -1.584  -0.812
 -1.585  -0.793
 -1.585  -0.774
 -1.585  -0.754
 -1.585  -0.735
 -1.585  -0.716
 -1.585  -0.697
 -1.584  -0.678
 -1.584  -0.659
 -1.584  -0.639
 -1.583  -0.620
 -1.583  -0.601
 -1.582  -0.581
 -1.581  -0.562
 -1.580  -0.543
 -1.579  -0.523
 -1.577  -0.504
 -1.575  -0.484
 -1.573  -0.464
 -1.570  -0.444
 -1.566  -0.424
 -1.562  -0.404
 -1.557  -0.384
 -1.550  -0.363
 -1.543  -0.342
 -1.533  -0.320
 -1.522  -0.298
 -1.507  -0.276
 -1.490  -0.252
 -1.468  -0.227
 -1.440  -0.201
 -1.404  -0.173
 -1.357  -0.142
 -1.293  -0.107
 -1.202  -0.065
 -1.061  -0.011
 -0.795   0.075
  2.449   0.905
/
\put {$m_\sigma=48$~GeV} [l] at -1.4 -1.5
\setdots <2pt>
\plot
 -2.000  -0.139
  0.100   0.386
  2.200   0.911
/
\setsolid
\put {$m_t=110$~GeV} at 1 -2.25
\plot
 -2.000  -3.618
 -1.790  -3.433
 -1.580  -3.254
 -1.370  -3.082
 -1.160  -2.916
 -0.950  -2.757
 -0.740  -2.605
 -0.530  -2.460
 -0.320  -2.320
 -0.110  -2.187
  0.100  -2.058
  0.310  -1.933
  0.520  -1.813
  0.730  -1.696
  0.940  -1.582
  1.150  -1.470
  1.360  -1.360
  1.570  -1.252
  1.780  -1.145
  1.990  -1.040
  2.200  -0.935
/
\setdashes
\put {$m_t=150$~GeV} at 1 -.75
\plot
 -2.000  -3.632
 -1.790  -3.440
 -1.580  -3.252
 -1.370  -3.070
 -1.160  -2.894
 -0.950  -2.725
 -0.740  -2.562
 -0.530  -2.406
 -0.320  -2.257
 -0.110  -2.114
  0.100  -1.976
  0.310  -1.844
  0.520  -1.716
  0.730  -1.592
  0.940  -1.473
  1.150  -1.356
  1.360  -1.242
  1.570  -1.130
  1.780  -1.021
  1.990  -0.913
  2.200  -0.806
/
\endpicture
$$
}
\caption{Parameter space excluded by $B^0$-$\overline{B^0}$ mixing; the region
above the B-line
and to the right of the 48 GeV contour is allowed.}
\label{fig:plot2}
\end{figure}

We must also consider the possibility of detecting the isotriplet pions
$\pi_p$. In fig.\hspace{1ex}\ref{fig:plot3}, we show the contour of the
allowed region from Fig.~\ref{fig:plot2} (with $m_t=100$~GeV), with the
$m_\pi=m_Z/2$,
and $m_\pi=m_W$ lines as points of reference. For all of the
parameter space above the B-line, $m_\pi>m_Z/2$ so that the decay
$Z\rightarrow 2\,\pi_p$ is not kinematically allowed. In particular,
we obtain little useful information from the available LEP limits on
$Z\rightarrow\pi_p^+\pi_p^-$. In fact, for most of the parameter
space, $m_\pi$ is larger than $m_Z$, and the triplet pions cannot be
produced in any weak gauge boson decays.

\begin{figure}[htb]
\finput{$$
\beginpicture
\setcoordinatesystem units <90\tdim,60\tdim>
\setplotarea x from -2 to 2.2, y from -5 to 1.1
\linethickness=0pt
\putrule from 0 -5.3 to 0 1.7
\inboundscheckon
\linethickness=.013truein
\axis bottom ticks long in logged at .01 .1 1 10 100 / /
\axis top ticks long in logged at .01 .1 1 10 100 / /
\put {$10^{-2}$} [b] at -2 -5.35
\put {$10^{-1}$} [b] at -1 -5.35
\put {$1$} [b] at -0 -5.35
\put {$2$} [b] at .3 -5.35
\put {$5$} [b] at .7 -5.35
\put {$10$} [b] at 1 -5.35
\put {$10^2$} [b] at 2 -5.35
\linethickness=.006truein
\axis bottom ticks short in logged unlabeled at .02 .2 2 20 / /
\axis bottom ticks short in logged unlabeled at .03 .3 3 30 / /
\axis bottom ticks short in logged unlabeled at .04 .4 4 40 / /
\axis bottom ticks short in logged unlabeled at .05 .5 5 50 / /
\axis top ticks short in logged unlabeled at .02 .2 2 20 / /
\axis top ticks short in logged unlabeled at .03 .3 3 30 / /
\axis top ticks short in logged unlabeled at .04 .4 4 40 / /
\axis top ticks short in logged unlabeled at .05 .5 5 50 / /
\linethickness=.013truein
\put {$10^{-5}$} [r] at -2.08 -5
\put {$10^{-4}$} [r] at -2.08 -4
\put {$10^{-3}$} [r] at -2.08 -3
\put {$10^{-2}$} [r] at -2.08 -2
\put {$10^{-1}$} [r] at -2.08 -1
\put {$1$} [r] at -2.08 -0
\put {$2$} [r] at -2.08 .3
\put {$5$} [r] at -2.08 .7
\put {$10$} [r] at -2.08 1
\axis right ticks long in logged unlabeled at .00001 .0001 .001 .01 .1 1  10 /
/
\axis left ticks long in logged unlabeled at .00001 .0001 .001 .01 .1 1 10 / /
\linethickness=.006truein
\axis left ticks short in logged unlabeled at .00002 .0002 .002 .02 .2 2 / /
\axis left ticks short in logged unlabeled at .00003 .0003 .003 .03 .3 3 / /
\axis left ticks short in logged unlabeled at .00004 .0004 .004 .04 .4 4 / /
\axis left ticks short in logged unlabeled at .00005 .0005 .005 .05 .5 5 / /
\axis right ticks short in logged unlabeled at .00002 .0002 .002 .02 .2 2 / /
\axis right ticks short in logged unlabeled at .00003 .0003 .003 .03 .3 3 / /
\axis right ticks short in logged unlabeled at .00004 .0004 .004 .04 .4 4 / /
\axis right ticks short in logged unlabeled at .00005 .0005 .005 .05 .5 5 / /
\put {$\tilde\lambda\rightarrow$} at .1 -5.6
\put {\stack{$\uparrow$,$h$}} at -2.5 -2
\plot
 -1.320  -3.043
 -1.328  -3.025
 -1.335  -3.008
 -1.343  -2.991
 -1.350  -2.974
 -1.357  -2.956
 -1.364  -2.939
 -1.371  -2.921
 -1.377  -2.904
 -1.384  -2.886
 -1.390  -2.869
 -1.396  -2.851
 -1.401  -2.833
 -1.407  -2.816
 -1.412  -2.798
 -1.418  -2.780
 -1.423  -2.762
 -1.428  -2.744
 -1.433  -2.726
 -1.437  -2.708
 -1.442  -2.690
 -1.446  -2.672
 -1.450  -2.654
 -1.455  -2.636
 -1.459  -2.618
 -1.463  -2.600
 -1.466  -2.582
 -1.470  -2.563
 -1.474  -2.545
 -1.477  -2.527
 -1.480  -2.509
 -1.484  -2.490
 -1.487  -2.472
 -1.490  -2.453
 -1.493  -2.435
 -1.496  -2.417
 -1.499  -2.398
 -1.501  -2.380
 -1.504  -2.361
 -1.507  -2.343
 -1.509  -2.324
 -1.512  -2.306
 -1.514  -2.287
 -1.516  -2.269
 -1.519  -2.250
 -1.521  -2.231
 -1.523  -2.213
 -1.525  -2.194
 -1.527  -2.176
 -1.529  -2.157
 -1.531  -2.138
 -1.532  -2.120
 -1.534  -2.101
 -1.536  -2.082
 -1.538  -2.063
 -1.539  -2.045
 -1.541  -2.026
 -1.542  -2.007
 -1.544  -1.988
 -1.545  -1.970
 -1.547  -1.951
 -1.548  -1.932
 -1.549  -1.913
 -1.551  -1.894
 -1.552  -1.876
 -1.553  -1.857
 -1.554  -1.838
 -1.555  -1.819
 -1.556  -1.800
 -1.557  -1.781
 -1.558  -1.762
 -1.559  -1.743
 -1.560  -1.725
 -1.561  -1.706
 -1.562  -1.687
 -1.563  -1.668
 -1.564  -1.649
 -1.565  -1.630
 -1.566  -1.611
 -1.566  -1.592
 -1.567  -1.573
 -1.568  -1.554
 -1.569  -1.535
 -1.569  -1.516
 -1.570  -1.497
 -1.571  -1.478
 -1.571  -1.459
 -1.572  -1.440
 -1.573  -1.421
 -1.573  -1.402
 -1.574  -1.383
 -1.574  -1.364
 -1.575  -1.345
 -1.575  -1.326
 -1.576  -1.307
 -1.576  -1.288
 -1.577  -1.269
 -1.577  -1.250
 -1.578  -1.231
 -1.578  -1.212
 -1.579  -1.193
 -1.579  -1.174
 -1.580  -1.155
 -1.580  -1.136
 -1.580  -1.117
 -1.581  -1.098
 -1.581  -1.079
 -1.581  -1.060
 -1.582  -1.041
 -1.582  -1.022
 -1.582  -1.003
 -1.583  -0.984
 -1.583  -0.965
 -1.583  -0.945
 -1.583  -0.926
 -1.584  -0.907
 -1.584  -0.888
 -1.584  -0.869
 -1.584  -0.850
 -1.584  -0.831
 -1.584  -0.812
 -1.585  -0.793
 -1.585  -0.774
 -1.585  -0.754
 -1.585  -0.735
 -1.585  -0.716
 -1.585  -0.697
 -1.584  -0.678
 -1.584  -0.659
 -1.584  -0.639
 -1.583  -0.620
 -1.583  -0.601
 -1.582  -0.581
 -1.581  -0.562
 -1.580  -0.543
 -1.579  -0.523
 -1.577  -0.504
 -1.575  -0.484
 -1.573  -0.464
 -1.570  -0.444
 -1.566  -0.424
 -1.562  -0.404
 -1.557  -0.384
 -1.550  -0.363
 -1.543  -0.342
 -1.533  -0.320
 -1.522  -0.298
 -1.507  -0.276
 -1.490  -0.252
 -1.468  -0.227
 -1.440  -0.201
 -1.404  -0.173
 -1.357  -0.142
 -1.293  -0.107
 -1.202  -0.065
 -1.061  -0.011
 -0.795   0.075
  2.449   0.905
/
\plot
 -1.320  -3.043
 -1.160  -2.916
 -0.950  -2.757
 -0.740  -2.605
 -0.530  -2.460
 -0.320  -2.320
 -0.110  -2.187
  0.100  -2.058
  0.310  -1.933
  0.520  -1.813
  0.730  -1.696
  0.940  -1.582
  1.150  -1.470
  1.360  -1.360
  1.570  -1.252
  1.780  -1.145
  1.990  -1.040
  2.200  -0.935
/
\setdots <2pt>
\plot
 -2.000  -0.139
  0.100   0.386
  2.200   0.911
/
\plot
 -2.000  -1.487
  0.100  -0.962
  2.200  -0.437
/
\put {$hf'=0.2f$} at .1 -.7
\setdashes <10\tdim>
\plot
 -2.000  -2.617
  0.100  -3.667
  2.200  -4.717
/
\put {$m_\pi=m_Z/2$} at 1 -4.5
\setdashes <5\tdim>
\plot
 -2.000  -1.885
  0.100  -2.935
  2.200  -3.985
/
\put {$m_\pi=m_W$} at 1 -3.1
\endpicture
$$
}
\caption{Parameter space showing the
$m_\pi=m_Z/2$, $m_\pi=m_W$ and $\grotz f'=0.2f$ lines.}
\label{fig:plot3}
\end{figure}

\section{Oblique Corrections}\label{sec:sandt}

For completeness, we now will consider the effect of the nonstandard scalars
in our model on oblique electroweak radiative corrections. In particular,
we will estimate the $S$ and $T$ parameters, which have
been shown to provide stringent constraints on simple technicolor
models~\cite{pesktak}. The low-energy contributions to these parameters
can be computed within the framework of our effective chiral lagrangian
by evaluating pion loop diagrams. The results for $S$ will
depend only on our location in the $\lambda$-$\grotz$ plane, and on the
unknown coefficients in the chiral lagrangian. To compute $T$,
however, we will also need to specify the parameter $\delta$,
which determines the magnitude of custodial isospin-violation in the
technicolor sector. Since we do not know the value of $\delta$, the
sign of the non-standard contribution, or size of the top-bottom mass
splitting, our ability to constraint the model is somewhat diminished.
What we will show is that the new contributions to $T$ are not dangerously
large in much of the parameter space, assuming the worst case scenario
in which $\delta=1$. Thus, we will demonstrate that
$\grotz_+$ and $\grotz_-$ do not require fine-tuning in this model. Note
that we have separated the discussion of oblique corrections from
Section~\ref{sec:phenom} because we will be unable to exclude any additional
regions of the $\lambda$-$\grotz$ plane.

The $S$ parameter can be expressed as
\beq
S = -16\pi\left[\frac{d}{dq^2} \Pi_{3B}\right]_{q^2=0}
\label{eq:sdef}
\eeq
where $\Pi_{3B}$ is the piece of $W_3$$B$ vacuum polarization
proportional to $ig^{\mu\nu}$. There is a tree-level contribution
to $S$ from the following term in the chiral expansion,
\beq
\frac{c_2}{16\pi^2}\mbox{Tr}(\Sigma^\dagger W^{\mu\nu}\Sigma
B_{\mu\nu})
\label{eq:trees}
\eeq
where $W^{\mu\nu}$ and $B^{\mu\nu}$ are the gauge field strength
tensors, and where $c_2$ is a constant of order unity. The contribution to
$S$ is then given by
\beq
S_0=\frac{1}{\pi}c_2\,.
\label{eq:szero}
\eeq
$S_0$ originates from the nonperturbative, high-energy
dynamics, and hence is proportional to an undetermined parameter
in the low-energy theory. This high-energy technicolor contribution
to $S$ has been estimated in ref.~\cite{pesktak}, and is given
by\footnote{In ref.\cite{carone} this estimate
was multiplied by an extra factor of $v^2/f^2$; we now
believe this is an error. Without this factor, the value of
$S_0$ in ref.~\cite{carone} is reduced and the conclusion that
the massive scalar model is viable is strengthened.}
\beq
S_0\approx .3\frac{N_{TF}}{2}\frac{N}{3} =0.1 N
\label{eq:szest}
\eeq
for a model with two techniflavors. In this estimate, $S$ is
normalized so that $S=0$ corresponds to the standard model with
a $1$ TeV Higgs. We adopt this convention throughout. Assuming that
$S$ is positive, then the experimental upper bound is given by~\cite{tak}
\beq
S<0.9
\label{eq:sexp}
\eeq
at the 95\% confidence level. We need to verify that the low-energy
contributions in our model do not exceed the difference between
(\ref{eq:sexp}) and (\ref{eq:szest}), for an appropriate choice of
$N$.

\begin{figure}[htb]
\finput{$$
\beginpicture
\setcoordinatesystem units <\tdim,\tdim>
\put{\beginpicture
\setcoordinatesystem units <\tdim,\tdim>
\stpltsmbl
\linethickness=0pt
\putrule from -110 0 to 110 0
\putrule from 0 -30 to 90 -30
\multiput {\copy\phru} at -90 0 *5 10 0 /
\multiput {\copy\phru} at 30 0 *5 10 0 /
\setdashes
\circulararc 360 degrees from 30 0 center at 0 0
\put {$W_3$} at -105 0
\put {$B$} at 100 0
\put {$\pi_p$} at 0 40
\put {$\pi_p$} at 0 -40
\endpicture} at -125 0
\put{\beginpicture
\setcoordinatesystem units <\tdim,\tdim>
\stpltsmbl
\linethickness=0pt
\putrule from -110 0 to 110 0
\putrule from 0 -30 to 90 -30
\multiput {\copy\phru} at -90 0 *5 10 0 /
\multiput {\copy\phru} at 30 0 *5 10 0 /
\circulararc 180 degrees from -30 0 center at 0 0
\setdashes
\circulararc 180 degrees from 30 0 center at 0 0
\put {$W_3$} at -105 0
\put {$B$} at 100 0
\put {$\pi_p$} at 0 40
\put {$\sigma$} at 0 -40
\endpicture} at 130 0
\put {(a)} at -122 -65
\put {(b)} at 133 -65
\put{\beginpicture
\setcoordinatesystem units <\tdim,\tdim>
\stpltsmbl
\linethickness=0pt
\putrule from -110 0 to 110 0
\putrule from 0 -30 to 90 -30
\multiput {\copy\phru} at -90 0 *5 10 0 /
\multiput {\copy\phru} at 30 0 *5 10 0 /
\circulararc 180 degrees from 30 0 center at 0 0
\setquadratic
\plot
-30.0  -0.0
-32.9  -3.0
-29.5  -5.5
-26.0  -7.4
-28.0 -10.8
-29.5 -14.7
-25.5 -15.8
-21.5 -16.3
-22.2 -20.2
-22.2 -24.4
-18.1 -23.9
-14.2 -23.0
-13.4 -26.9
-11.9 -30.8
 -8.2 -28.9
 -5.0 -26.5
 -2.8 -29.9
  0.0 -33.0
  2.8 -29.9
  5.0 -26.5
  8.2 -28.9
 11.9 -30.8
 13.4 -26.9
 14.2 -23.0
 18.1 -23.9
 22.2 -24.4
 22.2 -20.2
 21.5 -16.3
 25.5 -15.8
 29.5 -14.7
 28.0 -10.8
 26.0  -7.4
 29.5  -5.5
 32.9  -3.0
 30.0   0.0
/
\put {$W_3$} at -105 0
\put {$B$} at 100 0
\put {$Z$} at 0 -45
\put {$\sigma$} at 0 40
\endpicture} at 0 -140
\put {(c)} at 3 -210
\endpicture
$$
}
\caption{Vacuum polarization diagrams contributing to $\Pi_{3B}$.}
\label{fig:vacpol}
\end{figure}

The low-energy contributions to $\Pi_{3B}$ in our model are obtained
from the one-loop vacuum polarization diagrams shown in
figs.\hspace{1ex}\ref{fig:vacpol}a, b and c. Working in the
$\overline{{\rm MS}}$ prescription, we find
\beq
\Delta S_a = \frac{1}{12\pi}\log(\frac{\mu^2}{m_\pi^2})
\label{eq:dsa}
\eeq
\begin{eqnarray}
\lefteqn{\Delta S_b =} \nonumber \\
&&-\frac{1}{12\pi}\frac{f^2}{v^2}\left[\log (
\frac{\mu^2}{m_\pi^2})
-2(x+x^2)+(3x^2+2x^3)\log (1+\frac{1}{x})+\frac{5}{6}
\right]
\label{eq:dsb}
\end{eqnarray}
\begin{eqnarray}
\lefteqn{\Delta S_c =} \nonumber \\
&& -\frac{1}{12\pi}\frac{{f'}^2}{v^2}
\left[\log(\frac{\mu^2}{m_\sigma^2})
+(4y+10y^2)-(9y^2+10y^3)\log(1+\frac{1}{y})+\frac{5}{6}\right]
\label{eq:dsc}
\end{eqnarray}
where $x\equiv m_\sigma^2/(m_\pi^2-m_\sigma^2)$ and
$y\equiv m_Z^2/(m_\sigma^2-m_Z^2)$. Note that the
sum of the three diagrams that we consider here is finite,
and therefore the sum of (\ref{eq:dsa}), (\ref{eq:dsb}),
and (\ref{eq:dsc}) is independent of the cutoff $\mu$.
Since a reference value of $S$ has already been subtracted in the
definition of $S_0$, as we pointed out earlier, our final estimate of
the low-energy contribution is given by
\beq
\Delta S = \Delta S_a + \Delta S_b + \Delta S_c\,.
\eeq

To get some idea of the characteristic size of $\Delta S$, we evaluate it
along the $\grotz=0.2 f/f'$ line shown in fig.\hspace{1ex}\ref{fig:plot3}.
This line was chosen because it is in the heart of the allowed
region, and well within the area where our chiral lagrangian is valid.
Along this path, the value of $\Delta S$ is plotted
in fig.\hspace{1ex}\ref{fig:plot4}. We see that $\Delta S$ ranges from
roughly $-0.11$ to $0.01$; in any event, it is not very large.
For a reasonable choice of $N$, e.g. $N=4$, the high energy
contribution is 0.4, and dominates over the low-energy component. The
total contribution in this particular example then ranges from 0.29 to 0.41,
which is consistent with the experimental bounds.

\begin{figure}[htb]
\finput{$$
\beginpicture
\setcoordinatesystem units <90\tdim,900\tdim>
\setplotarea x from -1.523 to 2.2, y from -.2 to .1
\linethickness=0pt
\putrule from 0 -.22 to 0 .13
\inboundscheckon
\linethickness=.013truein
\axis bottom ticks long in logged at .1 1 10 100 / /
\axis top ticks long in logged at .1 1 10 100 / /
\put {$10^{-1}$} [b] at -1 -.225
\put {$1$} [b] at -0 -.225
\put {$2$} [b] at .3 -.225
\put {$5$} [b] at .7 -.225
\put {$10$} [b] at 1 -.225
\put {$10^2$} [b] at 2 -.225
\linethickness=.006truein
\axis bottom ticks short in logged unlabeled at .2 2 20 / /
\axis bottom ticks short in logged unlabeled at .03 .3 3 30 / /
\axis bottom ticks short in logged unlabeled at .04 .4 4 40 / /
\axis bottom ticks short in logged unlabeled at .05 .5 5 50 / /
\axis top ticks short in logged unlabeled at .2 2 20 / /
\axis top ticks short in logged unlabeled at .03 .3 3 30 / /
\axis top ticks short in logged unlabeled at .04 .4 4 40 / /
\axis top ticks short in logged unlabeled at .05 .5 5 50 / /
\linethickness=.013truein
\put {$-0.20$} [r] at -1.65 -.2
\put {$-0.15$} [r] at -1.65 -.15
\put {$-0.10$} [r] at -1.65 -.1
\put {$-0.05$} [r] at -1.65 -.05
\put {$0.00$} [r] at -1.65 0
\put {$0.05$} [r] at -1.65 .05
\put {$0.10$} [r] at -1.65 .1
\axis right ticks long in unlabeled at -.2 -.15 -.10 -.05 0 .05 .1 / /
\axis left ticks long in unlabeled at -.2 -.15 -.10 -.05 0 .05 .1 / /
\linethickness=.006truein
\axis right ticks short in unlabeled at -.19 -.14 -.09 -.04 0.01 .06 / / \axis
right ticks short in unlabeled at -.18 -.13 -.08 -.03 0.02 .07 / / \axis right
ticks short in unlabeled at -.17 -.12 -.07 -.02 0.03 .08 / / \axis right ticks
short in unlabeled at -.16 -.11 -.06 -.01 0.04 .09 / / \axis left ticks short
in unlabeled at -.19 -.14 -.09 -.04 0.01 .06 / / \axis left ticks short in
unlabeled at -.18 -.13 -.08 -.03 0.02 .07 / / \axis left ticks short in
unlabeled at -.17 -.12 -.07 -.02 0.03 .08 / / \axis left ticks short in
unlabeled at -.16 -.11 -.06 -.01 0.04 .09 / / \put
{$\tilde\lambda\rightarrow$} at .35 -.24
\put {\stack{$\uparrow$,$\Delta S$}} at -2.35 -.05
\setquadratic
\plot
-1.523
 -0.114
-1.301
 -0.102
-1.155
 -0.094
-1.000
 -0.085
-0.301
 -0.048
-0.046
 -0.037
 0.477
 -0.020
 0.845
 -0.011
 1.079
 -0.006
 1.716
  0.003
 2.182
  0.008
/
\put {$\Delta S$ along $hf'=0.2f$} at .45 .05
\endpicture
$$
}
\caption{Low-energy contribution to S.}
\label{fig:plot4}
\end{figure}

The $T$ parameter is defined in terms of a different combination of
vacuum polarizations
\beq
T=\frac{4\pi}{s^2c^2m_Z^2}\left[\Pi_{11}-\Pi_{33}\right]
_{q^2=0}\,.
\label{eq:tdef}
\eeq
To estimate $T$, we again start by considering the tree-level contributions
to $\Pi_{11}-\Pi_{33}$, which come from terms in our chiral expansion of the
form
\beq
{\cal L}=\frac{c_3}{16\pi^2}
\left({\rm Tr} \Phi\Grotz D^\mu \Sigma ^\dagger
\right)^2\,.
\label{eq:ttree}
\eeq
{}From (\ref{eq:ttree}) we obtain a result of order
\beq
\Delta T = \frac{1}{8 \pi^2 \alpha}\left( \frac{f'}{v}\right)^2
\grotz^2 \delta^2
\label{eq:dttree}
\eeq
which is relatively small throughout much of the allowed region.
However, (\ref{eq:dttree}) does not represent the leading contribution.
The one-loop diagrams that incorporate the mass splitting
of the pion triplet give us contributions to $T$ that are logarithmically
enhanced over (\ref{eq:dttree}). The pion masses are
split at ${\cal O}(\Grotz^2)$ by the effects of the term
\beq
c_4 f^2 {\rm Tr} (\Phi\Grotz\Sigma^\dagger
\Phi\Grotz\Sigma^\dagger) + \mbox{h.c..}
\label{eq:splitterm}
\eeq
{}From (\ref{eq:splitterm}) we find
\beq
m^2_{\pi^0}-m^2_{\pi^+} = 4 c_4 v^2 \grotz^2 \delta^2\,.
\label{eq:msplit}
\eeq
\begin{figure}[htb]
\finput{$$
\beginpicture
\setcoordinatesystem units <\tdim,\tdim>
\put{\beginpicture
\setcoordinatesystem units <\tdim,\tdim>
\stpltsmbl
\linethickness=0pt
\putrule from -110 0 to 110 0
\putrule from 0 -30 to 90 -30
\multiput {\copy\phru} at -90 0 *5 10 0 /
\multiput {\copy\phru} at 30 0 *5 10 0 /
\setdashes
\circulararc 360 degrees from 30 0 center at 0 0
\put {$W_1$} at -105 0
\put {$W_1$} at 105 0
\put {$\pi_p$} at 0 40
\put {$\pi_p$} at 0 -40
\endpicture} at -145 50
\put{\beginpicture
\setcoordinatesystem units <\tdim,\tdim>
\stpltsmbl
\linethickness=0pt
\putrule from -110 0 to 110 0
\putrule from 0 -30 to 90 -30
\multiput {\copy\phru} at -90 0 *5 10 0 /
\multiput {\copy\phru} at 30 0 *5 10 0 /
\circulararc 180 degrees from -30 0 center at 0 0
\setdashes
\circulararc 180 degrees from 30 0 center at 0 0
\put {$W_1$} at -105 0
\put {$W_1$} at 105 0
\put {$\pi_p$} at 0 40
\put {$\sigma$} at 0 -40
\endpicture} at -145 -50
\put{\beginpicture
\setcoordinatesystem units <\tdim,\tdim>
\stpltsmbl
\linethickness=0pt
\putrule from -110 0 to 110 0
\putrule from 0 -30 to 90 -30
\multiput {\copy\phru} at -90 0 *5 10 0 /
\multiput {\copy\phru} at 30 0 *5 10 0 /
\setdashes
\circulararc 360 degrees from 30 0 center at 0 0
\put {$W_3$} at -105 0
\put {$W_3$} at 105 0
\put {$\pi_p$} at 0 40
\put {$\pi_p$} at 0 -40
\endpicture} at 145 50
\put{\beginpicture
\setcoordinatesystem units <\tdim,\tdim>
\stpltsmbl
\linethickness=0pt
\putrule from -110 0 to 110 0
\putrule from 0 -30 to 90 -30
\multiput {\copy\phru} at -90 0 *5 10 0 /
\multiput {\copy\phru} at 30 0 *5 10 0 /
\circulararc 180 degrees from -30 0 center at 0 0
\setdashes
\circulararc 180 degrees from 30 0 center at 0 0
\put {$W_3$} at -105 0
\put {$W_3$} at 105 0
\put {$\pi_p$} at 0 40
\put {$\sigma$} at 0 -40
\endpicture} at 145 -50
\linethickness=1.5\tdim
\putrule from -270 100 to -270 -100
\putrule from -270 100 to -265 100
\putrule from -270 -100 to -265 -100
\putrule from -20 100 to -20 -100
\putrule from -20 100 to -25 100
\putrule from -20 -100 to -25 -100
\putrule from 270 100 to 270 -100
\putrule from 270 100 to 265 100
\putrule from 270 -100 to 265 -100
\putrule from 20 100 to 20 -100
\putrule from 20 100 to 25 100
\putrule from 20 -100 to 25 -100
\put {$-$} at 0 0
\endpicture
$$
}
\caption{Vacuum polarization diagrams contributing to $\Pi_{11}-\Pi_{33}$.}
\label{fig:tfeyn}
\end{figure}
We may now compute the one-loop
diagrams shown in Figure~\ref{fig:tfeyn}, by working to lowest order in
$m^2_{\pi^0}-m^2_{\pi^+}$, and substituting (\ref{eq:msplit}).
We obtain
\beq
\Delta T = \frac{4 c_4}{\alpha} \grotz^2 \delta^2
\left[ \left(\frac{f}{v}\right)^2 F(m_\sigma, m_{\pi^0})
+  F(m_{\pi^+}, m_{\pi^0})\right]
\label{eq:onelpt}
\eeq
where
\beq
F(m_1\,,\,m_2)=\frac{1}{16\pi^2}\left[
\log(\frac{\mu^2}{m_2^2})+\frac{m_1^4}
{(m_1^2-m_2^2)^2}\log(\frac{m_2^2}{m_1^2})
+\frac{m_1^2}{(m_1^2-m_2^2)}\right]
\label{eq:func}
\eeq
and where we set $\mu$ equal to the technicolor scale, $4\pi f$.
In fact, only the logarithmically enhanced term is of
any real interest to us in this result. The $1/\epsilon$ poles in the
loop integrals have been removed by introducing counterterms of the
form (\ref{eq:ttree}). However, as we have seen, these terms also
contribute an unknown finite amount to $\Delta T$. The point of our
computation is that the logarithmic term is large enough so that we don't
need to worry about all the unknown finite contributions in order to
obtain a reliable estimate.

Again we evaluate our results along the $\grotz=0.2 f/f'$ line shown in
fig.\hspace{1ex}\ref{fig:plot3}, setting $\delta=1$.
The results are plotted in fig.\hspace{1ex}\ref{fig:t}.
\begin{figure}[htb]
\finput{$$
\beginpicture
\setcoordinatesystem units <90\tdim,300\tdim>
\setplotarea x from -1.523 to 2.2, y from 0 to .8
\linethickness=0pt
\putrule from 0 -.1 to 0 1
\inboundscheckon
\linethickness=.013truein
\axis bottom ticks long in logged at .1 1 10 100 / /
\axis top ticks long in logged at .1 1 10 100 / /
\put {$10^{-1}$} [b] at -1 -.08
\put {$1$} [b] at -0 -.08
\put {$2$} [b] at .3 -.08
\put {$5$} [b] at .7 -.08
\put {$10$} [b] at 1 -.08
\put {$10^2$} [b] at 2 -.08
\linethickness=.006truein
\axis bottom ticks short in logged unlabeled at .2 2 20 / /
\axis bottom ticks short in logged unlabeled at .03 .3 3 30 / /
\axis bottom ticks short in logged unlabeled at .04 .4 4 40 / /
\axis bottom ticks short in logged unlabeled at .05 .5 5 50 / /
\axis top ticks short in logged unlabeled at .2 2 20 / /
\axis top ticks short in logged unlabeled at .03 .3 3 30 / /
\axis top ticks short in logged unlabeled at .04 .4 4 40 / /
\axis top ticks short in logged unlabeled at .05 .5 5 50 / /
\linethickness=.013truein
\put {$0.0$} [r] at -1.65 0.0
\put {$0.2$} [r] at -1.65 0.2
\put {$0.4$} [r] at -1.65 0.4
\put {$0.6$} [r] at -1.65 0.6
\put {$0.8$} [r] at -1.65 0.8
\axis right ticks long in unlabeled at 0 .2 .4 .6 .8 / /
\axis left ticks long in unlabeled at 0 .2 .4 .6 .8 / /
\linethickness=.006truein
\axis right ticks long in unlabeled at 0.05 .25 .45 .65 / /
\axis right ticks long in unlabeled at 0.15 .35 .55 .75 / /
\axis right ticks long in unlabeled at 0.1 .3 .5 .7 / /
\axis left ticks long in unlabeled at 0.05 .25 .45 .65 / /
\axis left ticks long in unlabeled at 0.15 .35 .55 .75 / /
\axis left ticks long in unlabeled at 0.1 .3 .5 .7 / /
\put {$\tilde\lambda\rightarrow$} at .35 -.15
\put {\stack{$\uparrow$,$\Delta T$}} at -2.15 .4
\setquadratic
\plot
-1.523
  0.017
-1.301
  0.024
-1.155
  0.027
-1.046
  0.030
-1.000
  0.032
-0.523
  0.057
-0.301
  0.075
-0.155
  0.089
-0.046
  0.102
 0.000
  0.107
 0.477
  0.188
 0.699
  0.242
 0.845
  0.283
 0.954
  0.318
 1.079
  0.360
 1.716
  0.623
 2.182
  0.789
/
\put {$\Delta T$ along $hf'=0.2f$} at .45 .7
\put {$\delta=1$} at .45 .6
\endpicture
$$
}
\caption{Low-energy contribution to T.}
\label{fig:t}
\end{figure}

We see that these results present no immanent danger of conflict with the
experimental bounds on $T$, at least in most of the region of the parameter
space that we have sampled. In fact, we could adjust $\delta$ from $1$ to
$1/3$
and suppress the results shown by a factor of $1/9$, without introducing
any substantial fine-tuning. Since $\Delta T$ is proportional to
$\grotz^2$, it seems that the necessity of fine-tuning $\delta$ may arise
at larger values of $\grotz$, but this is exactly the limit in which our
chiral expansion begins to break down. We will therefore need
to reexamine this question in next section.

\section{Beyond the Chiral Expansion}\label{sec:beyond}

At the bottom of the allowed region in fig.\hspace{1ex}\ref{fig:plot3}, our
model looks like a rather standard technicolor model, like that discussed
in~\cite{carone}. However, near the top of the allowed region, the physics is
very different. A full discussion of the phenomenology of this region is
beyond the scope of this paper, but we would like at least to explain what is
so peculiar about it. The essential point is illustrated in
fig.\hspace{1ex}\ref{fig:fscale}, where we have ploted contours of fixed $f$
(the dashed lines) and $\grotz f'$ (the dotted lines).
\begin{figure}[htb]
\finput{$$
\beginpicture
\setcoordinatesystem units <90\tdim,60\tdim>
\setplotarea x from -2 to 2.2, y from -5 to 1.1
\linethickness=0pt
\putrule from 0 -5.3 to 0 1.7
\inboundscheckon
\linethickness=.013truein
\axis bottom ticks long in logged at .01 .1 1 10 100 / /
\axis top ticks long in logged at .01 .1 1 10 100 / /
\put {$10^{-2}$} [b] at -2 -5.35
\put {$10^{-1}$} [b] at -1 -5.35
\put {$1$} [b] at -0 -5.35
\put {$2$} [b] at .3 -5.35
\put {$5$} [b] at .7 -5.35
\put {$10$} [b] at 1 -5.35
\put {$10^2$} [b] at 2 -5.35
\linethickness=.006truein
\axis bottom ticks short in logged unlabeled at .02 .2 2 20 / /
\axis bottom ticks short in logged unlabeled at .03 .3 3 30 / /
\axis bottom ticks short in logged unlabeled at .04 .4 4 40 / /
\axis bottom ticks short in logged unlabeled at .05 .5 5 50 / /
\axis top ticks short in logged unlabeled at .02 .2 2 20 / /
\axis top ticks short in logged unlabeled at .03 .3 3 30 / /
\axis top ticks short in logged unlabeled at .04 .4 4 40 / /
\axis top ticks short in logged unlabeled at .05 .5 5 50 / /
\linethickness=.013truein
\put {$10^{-5}$} [r] at -2.08 -5
\put {$10^{-4}$} [r] at -2.08 -4
\put {$10^{-3}$} [r] at -2.08 -3
\put {$10^{-2}$} [r] at -2.08 -2
\put {$10^{-1}$} [r] at -2.08 -1
\put {$1$} [r] at -2.08 -0
\put {$2$} [r] at -2.08 .3
\put {$5$} [r] at -2.08 .7
\put {$10$} [r] at -2.08 1
\axis right ticks long in logged unlabeled at .00001 .0001 .001 .01 .1 1  10 /
/
\axis left ticks long in logged unlabeled at .00001 .0001 .001 .01 .1 1 10 / /
\linethickness=.006truein
\axis left ticks short in logged unlabeled at .00002 .0002 .002 .02 .2 2 / /
\axis left ticks short in logged unlabeled at .00003 .0003 .003 .03 .3 3 / /
\axis left ticks short in logged unlabeled at .00004 .0004 .004 .04 .4 4 / /
\axis left ticks short in logged unlabeled at .00005 .0005 .005 .05 .5 5 / /
\axis right ticks short in logged unlabeled at .00002 .0002 .002 .02 .2 2 / /
\axis right ticks short in logged unlabeled at .00003 .0003 .003 .03 .3 3 / /
\axis right ticks short in logged unlabeled at .00004 .0004 .004 .04 .4 4 / /
\axis right ticks short in logged unlabeled at .00005 .0005 .005 .05 .5 5 / /
\put {$\tilde\lambda\rightarrow$} at .1 -5.6
\put {\stack{$\uparrow$,$h$}} at -2.5 -2
\plot
 -1.320  -3.043
 -1.328  -3.025
 -1.335  -3.008
 -1.343  -2.991
 -1.350  -2.974
 -1.357  -2.956
 -1.364  -2.939
 -1.371  -2.921
 -1.377  -2.904
 -1.384  -2.886
 -1.390  -2.869
 -1.396  -2.851
 -1.401  -2.833
 -1.407  -2.816
 -1.412  -2.798
 -1.418  -2.780
 -1.423  -2.762
 -1.428  -2.744
 -1.433  -2.726
 -1.437  -2.708
 -1.442  -2.690
 -1.446  -2.672
 -1.450  -2.654
 -1.455  -2.636
 -1.459  -2.618
 -1.463  -2.600
 -1.466  -2.582
 -1.470  -2.563
 -1.474  -2.545
 -1.477  -2.527
 -1.480  -2.509
 -1.484  -2.490
 -1.487  -2.472
 -1.490  -2.453
 -1.493  -2.435
 -1.496  -2.417
 -1.499  -2.398
 -1.501  -2.380
 -1.504  -2.361
 -1.507  -2.343
 -1.509  -2.324
 -1.512  -2.306
 -1.514  -2.287
 -1.516  -2.269
 -1.519  -2.250
 -1.521  -2.231
 -1.523  -2.213
 -1.525  -2.194
 -1.527  -2.176
 -1.529  -2.157
 -1.531  -2.138
 -1.532  -2.120
 -1.534  -2.101
 -1.536  -2.082
 -1.538  -2.063
 -1.539  -2.045
 -1.541  -2.026
 -1.542  -2.007
 -1.544  -1.988
 -1.545  -1.970
 -1.547  -1.951
 -1.548  -1.932
 -1.549  -1.913
 -1.551  -1.894
 -1.552  -1.876
 -1.553  -1.857
 -1.554  -1.838
 -1.555  -1.819
 -1.556  -1.800
 -1.557  -1.781
 -1.558  -1.762
 -1.559  -1.743
 -1.560  -1.725
 -1.561  -1.706
 -1.562  -1.687
 -1.563  -1.668
 -1.564  -1.649
 -1.565  -1.630
 -1.566  -1.611
 -1.566  -1.592
 -1.567  -1.573
 -1.568  -1.554
 -1.569  -1.535
 -1.569  -1.516
 -1.570  -1.497
 -1.571  -1.478
 -1.571  -1.459
 -1.572  -1.440
 -1.573  -1.421
 -1.573  -1.402
 -1.574  -1.383
 -1.574  -1.364
 -1.575  -1.345
 -1.575  -1.326
 -1.576  -1.307
 -1.576  -1.288
 -1.577  -1.269
 -1.577  -1.250
 -1.578  -1.231
 -1.578  -1.212
 -1.579  -1.193
 -1.579  -1.174
 -1.580  -1.155
 -1.580  -1.136
 -1.580  -1.117
 -1.581  -1.098
 -1.581  -1.079
 -1.581  -1.060
 -1.582  -1.041
 -1.582  -1.022
 -1.582  -1.003
 -1.583  -0.984
 -1.583  -0.965
 -1.583  -0.945
 -1.583  -0.926
 -1.584  -0.907
 -1.584  -0.888
 -1.584  -0.869
 -1.584  -0.850
 -1.584  -0.831
 -1.584  -0.812
 -1.585  -0.793
 -1.585  -0.774
 -1.585  -0.754
 -1.585  -0.735
 -1.585  -0.716
 -1.585  -0.697
 -1.584  -0.678
 -1.584  -0.659
 -1.584  -0.639
 -1.583  -0.620
 -1.583  -0.601
 -1.582  -0.581
 -1.581  -0.562
 -1.580  -0.543
 -1.579  -0.523
 -1.577  -0.504
 -1.575  -0.484
 -1.573  -0.464
 -1.570  -0.444
 -1.566  -0.424
 -1.562  -0.404
 -1.557  -0.384
 -1.550  -0.363
 -1.543  -0.342
 -1.533  -0.320
 -1.522  -0.298
 -1.507  -0.276
 -1.490  -0.252
 -1.468  -0.227
 -1.440  -0.201
 -1.404  -0.173
 -1.357  -0.142
 -1.293  -0.107
 -1.202  -0.065
 -1.061  -0.011
 -0.795   0.075
  2.449   0.905
/
\plot
 -1.320  -3.043
 -1.160  -2.916
 -0.950  -2.757
 -0.740  -2.605
 -0.530  -2.460
 -0.320  -2.320
 -0.110  -2.187
  0.100  -2.058
  0.310  -1.933
  0.520  -1.813
  0.730  -1.696
  0.940  -1.582
  1.150  -1.470
  1.360  -1.360
  1.570  -1.252
  1.780  -1.145
  1.990  -1.040
  2.200  -0.935
/
\setdashes
\put {$25$} [l] at -1.8 -.86
\plot
 -2.000  -0.858
  0.100   1.242
  2.200   3.342
/
\put {$50$} [l] at -1.8 -1.78
\plot
 -2.000  -1.781
  0.100   0.319
  2.200   2.419
/
\put {$100$} [l] at -1.8 -2.77
\plot
 -2.000  -2.772
  0.100  -0.672
  2.200   1.428
/
\put {$150$} [l] at -1.8 -3.48
\plot
 -2.000  -3.477
  0.100  -1.377
  2.200   0.723
/
\put {$f=200$} [l] at -1.8 -4.23
\plot
 -2.000  -4.227
  0.100  -2.127
  2.200  -0.027
/
\setdots <3pt>
\put {$hf'=5$} [r] at 2 -1.45
\plot
  4.018  -0.731
  1.877  -1.246
  1.311  -1.369
  0.970  -1.437
  0.725  -1.481
  0.532  -1.514
  0.373  -1.538
  0.238  -1.558
  0.120  -1.573
  0.015  -1.586
 -0.079  -1.597
 -0.166  -1.606
 -0.245  -1.614
 -0.318  -1.621
 -0.386  -1.627
 -0.450  -1.633
 -0.511  -1.637
 -0.568  -1.642
 -0.622  -1.645
 -0.673  -1.649
 -0.722  -1.652
 -0.769  -1.655
 -0.814  -1.657
 -0.857  -1.659
 -0.898  -1.662
 -0.938  -1.664
 -0.977  -1.665
 -1.014  -1.667
 -1.050  -1.669
 -1.085  -1.670
 -1.119  -1.671
 -1.152  -1.673
 -1.184  -1.674
 -1.216  -1.675
 -1.246  -1.676
 -1.276  -1.677
 -1.304  -1.678
 -1.333  -1.678
 -1.360  -1.679
 -1.387  -1.680
 -1.413  -1.681
 -1.439  -1.681
 -1.464  -1.682
 -1.489  -1.683
 -1.513  -1.683
 -1.537  -1.684
 -1.560  -1.684
 -1.583  -1.685
 -1.605  -1.685
 -1.627  -1.686
 -1.648  -1.686
 -1.670  -1.686
 -1.690  -1.687
 -1.711  -1.687
 -1.731  -1.687
 -1.751  -1.688
 -1.770  -1.688
 -1.789  -1.688
 -1.808  -1.689
 -1.827  -1.689
 -1.845  -1.689
 -1.863  -1.689
 -1.881  -1.690
 -1.899  -1.690
 -1.916  -1.690
 -1.933  -1.690
 -1.950  -1.691
 -1.966  -1.691
 -1.983  -1.691
 -1.999  -1.691
 -2.015  -1.691
 -2.031  -1.692
 -2.046  -1.692
 -2.062  -1.692
 -2.077  -1.692
 -2.092  -1.692
 -2.107  -1.692
 -2.121  -1.693
 -2.136  -1.693
 -2.150  -1.693
 -2.164  -1.693
 -2.178  -1.693
 -2.192  -1.693
 -2.206  -1.693
 -2.219  -1.693
/
\put {$25$} [r] at 2 -.85
\plot
  3.795  -0.259
  1.941  -0.684
  1.388  -0.789
  1.048  -0.843
  0.801  -0.877
  0.605  -0.900
  0.442  -0.917
  0.302  -0.929
  0.179  -0.939
  0.069  -0.947
 -0.031  -0.953
 -0.122  -0.959
 -0.206  -0.963
 -0.284  -0.967
 -0.357  -0.970
 -0.425  -0.973
 -0.490  -0.975
 -0.551  -0.977
 -0.610  -0.979
 -0.665  -0.980
 -0.718  -0.982
 -0.769  -0.983
 -0.817  -0.984
 -0.864  -0.985
 -0.909  -0.986
 -0.952  -0.987
 -0.994  -0.988
 -1.035  -0.989
 -1.074  -0.989
 -1.112  -0.990
 -1.148  -0.990
 -1.184  -0.991
 -1.219  -0.991
 -1.253  -0.992
 -1.286  -0.992
 -1.318  -0.992
 -1.349  -0.993
 -1.380  -0.993
 -1.409  -0.993
 -1.438  -0.994
 -1.467  -0.994
 -1.495  -0.994
 -1.522  -0.994
 -1.548  -0.995
 -1.575  -0.995
 -1.600  -0.995
 -1.625  -0.995
 -1.650  -0.995
 -1.674  -0.996
 -1.697  -0.996
 -1.721  -0.996
 -1.743  -0.996
 -1.766  -0.996
 -1.788  -0.996
 -1.809  -0.996
 -1.831  -0.997
 -1.852  -0.997
 -1.872  -0.997
 -1.892  -0.997
 -1.912  -0.997
 -1.932  -0.997
 -1.951  -0.997
 -1.970  -0.997
 -1.989  -0.997
 -2.008  -0.997
 -2.026  -0.997
 -2.044  -0.997
 -2.062  -0.998
 -2.079  -0.998
 -2.096  -0.998
 -2.113  -0.998
 -2.130  -0.998
 -2.147  -0.998
 -2.163  -0.998
 -2.179  -0.998
 -2.195  -0.998
 -2.211  -0.998
 -2.227  -0.998
 -2.242  -0.998
 -2.257  -0.998
 -2.272  -0.998
 -2.287  -0.998
 -2.302  -0.998
 -2.316  -0.998
 -2.331  -0.998
 -2.345  -0.998
 -2.359  -0.998
 -2.373  -0.998
 -2.387  -0.999
 -2.400  -0.999
 -2.414  -0.999
 -2.427  -0.999
 -2.440  -0.999
/
\put {$100$} [r] at 2 -.36
\plot
  3.004   0.016
  2.094  -0.168
  1.654  -0.240
  1.358  -0.280
  1.132  -0.306
  0.949  -0.323
  0.794  -0.336
  0.659  -0.346
  0.539  -0.353
  0.432  -0.359
  0.334  -0.364
  0.244  -0.368
  0.160  -0.371
  0.082  -0.374
  0.010  -0.376
 -0.059  -0.378
 -0.124  -0.380
 -0.185  -0.381
 -0.244  -0.383
 -0.299  -0.384
 -0.353  -0.385
 -0.404  -0.386
 -0.453  -0.387
 -0.500  -0.388
 -0.545  -0.388
 -0.589  -0.389
 -0.631  -0.389
 -0.672  -0.390
 -0.712  -0.390
 -0.750  -0.391
 -0.787  -0.391
 -0.823  -0.392
 -0.858  -0.392
 -0.893  -0.392
 -0.926  -0.392
 -0.958  -0.393
 -0.990  -0.393
 -1.021  -0.393
 -1.051  -0.393
 -1.080  -0.394
 -1.109  -0.394
 -1.137  -0.394
 -1.164  -0.394
 -1.191  -0.394
 -1.218  -0.394
 -1.244  -0.395
 -1.269  -0.395
 -1.294  -0.395
 -1.318  -0.395
 -1.342  -0.395
 -1.365  -0.395
 -1.388  -0.395
 -1.411  -0.395
 -1.433  -0.395
 -1.455  -0.395
 -1.477  -0.396
 -1.498  -0.396
 -1.519  -0.396
 -1.539  -0.396
 -1.559  -0.396
 -1.579  -0.396
 -1.599  -0.396
 -1.618  -0.396
 -1.637  -0.396
 -1.655  -0.396
 -1.674  -0.396
 -1.692  -0.396
 -1.710  -0.396
 -1.728  -0.396
 -1.745  -0.396
 -1.762  -0.396
 -1.779  -0.396
 -1.796  -0.396
 -1.812  -0.397
 -1.829  -0.397
 -1.845  -0.397
 -1.861  -0.397
 -1.876  -0.397
 -1.892  -0.397
 -1.907  -0.397
 -1.923  -0.397
 -1.938  -0.397
 -1.952  -0.397
 -1.967  -0.397
 -1.981  -0.397
 -1.996  -0.397
 -2.010  -0.397
 -2.024  -0.397
 -2.038  -0.397
 -2.052  -0.397
 -2.065  -0.397
 -2.079  -0.397
 -2.092  -0.397
 -2.105  -0.397
 -2.118  -0.397
 -2.131  -0.397
/
\put {$400$} [r] at 2 .18
\plot
  2.950   0.481
  1.938   0.319
  1.461   0.271
  1.137   0.249
  0.889   0.236
  0.686   0.228
  0.513   0.223
  0.362   0.219
  0.228   0.217
  0.108   0.215
 -0.002   0.213
 -0.103   0.212
 -0.196   0.211
 -0.283   0.210
 -0.364   0.209
 -0.441   0.209
 -0.513   0.208
 -0.581   0.208
 -0.645   0.208
 -0.707   0.207
 -0.766   0.207
 -0.822   0.207
 -0.876   0.207
 -0.927   0.206
 -0.977   0.206
 -1.025   0.206
 -1.071   0.206
 -1.115   0.206
 -1.158   0.206
 -1.200   0.206
 -1.240   0.205
 -1.279   0.205
 -1.317   0.205
 -1.354   0.205
 -1.390   0.205
 -1.425   0.205
 -1.459   0.205
 -1.492   0.205
 -1.524   0.205
 -1.556   0.205
 -1.587   0.205
 -1.617   0.205
 -1.646   0.205
 -1.675   0.205
 -1.703   0.205
 -1.730   0.205
 -1.757   0.205
 -1.784   0.205
 -1.809   0.205
 -1.835   0.205
 -1.860   0.205
 -1.884   0.205
 -1.908   0.205
 -1.931   0.205
 -1.954   0.205
 -1.977   0.205
 -1.999   0.205
 -2.021   0.205
 -2.043   0.205
 -2.064   0.205
 -2.085   0.204
 -2.106   0.204
 -2.126   0.204
 -2.146   0.204
 -2.165   0.204
 -2.185   0.204
 -2.204   0.204
 -2.222   0.204
 -2.241   0.204
 -2.259   0.204
 -2.277   0.204
 -2.295   0.204
 -2.312   0.204
 -2.330   0.204
 -2.347   0.204
 -2.363   0.204
 -2.380   0.204
 -2.396   0.204
 -2.413   0.204
 -2.429   0.204
 -2.444   0.204
 -2.460   0.204
 -2.475   0.204
 -2.491   0.204
 -2.506   0.204
 -2.521   0.204
 -2.535   0.204
 -2.550   0.204
 -2.564   0.204
 -2.579   0.204
 -2.593   0.204
 -2.607   0.204
 -2.620   0.204
 -2.634   0.204
 -2.648   0.204
 -2.661   0.204
 -2.674   0.204
 -2.687   0.204
 -2.700   0.204
 -2.713   0.204
 -2.726   0.204
/
\endpicture
$$
}
\caption{Values of $f$ and $\grotz f'$.}
\label{fig:fscale}
\end{figure}

At the top of the allowed region, the technicolor scale, $f$, can be
considerably smaller than the average ``current'' mass of the technifermions
$\grotz_\pm f'$. While the ratio, $\grotz f'/f$ cannot grow arbitrarily large
because of the Coleman-Weinberg corrections discussed in Section~\ref{sec:cw},
the ratio does get close to $4\pi$, and that is enough to drastically change
the physics of the technifermion bound states. The technifermions in this
region are much more analogous to $c$ quarks in QCD than they are to $u$ and
$d$ quarks. In particular, one might expect that the vector bounds states of
these technifermions will be quite narrow, like the $J/\psi$.

Note also that $\grotz f'$ is the average current mass of the technifermions.
If $\delta$ is very different from zero, one of technifermion masses will be
smaller. This is quite interesting in the upper left-hand corner of the
allowed region, where the technicolor scale is quite small. Here one can
imagine that some of the technifermions bound states might even be light
enough to show up at LEPII energies. The lightest state will presumably be a
bound state of the lighter technifermions, and therefore it will be neutral
and can be produced at LEPII if it is light enough. On the other hand, this
may make a large contribution to the $T$ parameter through mixing of the
lightest state with the $Z$.

\section{Conclusions}\label{sec:conclude}

We have presented a minimal technicolor model, in which the
ordinary fermions and technifermions are coupled by a massless
scalar doublet. The technicolor condensate that breaks the
electroweak symmetry also drives the scalar's vacuum expectation value,
which in turn generates the fermion masses. Flavor symmetry breaking
originates from Yukawa couplings, as in the standard model. We have
shown that the scalar states in the theory can be made massive enough
to avoid detection, even though the high-energy theory has no mass terms,
and no arbitrary dimensionful parameters. In addition, we have shown
that the model does not generate unacceptably large flavor changing neutral
currents, nor does it give a large low-energy contribution to the
$S$ or $T$ parameter.

Our model is nearly as parsimonious as the standard model with a single
fundamental scalar doublet. There are only two more continuous parameters,
$\grotz_\pm$. The technicolor scale, $\Lambda_{TC}$, plays the role in our
model that the scalar mass term plays in the simplest standard model --- it
sets the mass scale for all light particle states. In terms of parameter
counting, {\bf our model is the simplest extension of the standard model}. The
next simplest, the two-doublet model, even with the scalar masses set to zero,
has one more renormalized parameter than ours.

In addition to aesthetics, there are also deep theoretical
reasons for favoring the model we have presented over the
version with $M_\phi > 0$. Let us consider the possibility that
the values of the fundamental constants of nature are inextricably
linked to the physics of wormholes~\cite{coleman1}. It has
been argued that the mechanism proposed by Coleman to
eliminate the cosmological constant~\cite{coleman2} also
eliminates adjustable scalar mass terms in the low-energy
theory~\cite{preskill}. Specifically, Preskill has shown that if scalars
or fermions have bare mass terms in the full theory above the wormhole
scale, then these states will have physical masses in the low-energy
theory that are either not much smaller than the wormhole scale,
or identically zero~\cite{preskill}. If this scenario is correct,
then electroweak symmetry breaking cannot be generated by the value
of a scalar mass term, but instead, must be dynamical in nature.
Although this argument is somewhat speculative, it does provide
a thought-provoking reason for studying models of this type.

\vspace{36pt}
\centerline{\bf Acknowledgments}

This work was supported in part by the National Science Foundation,
under grant \# PHY-9218167, and in part by the Texas National
Research Laboratory Commission under grant \# RGFY93-278B.

\appendix
\section{Appendix}

Below are the integrals relevant to the box diagram calculations
in Section~\ref{sec:fcnc}. They are in agreement with those published
previously in ref.~\cite{wise}.

\beq
I_1(m_1,m_2) = \int_E \frac{d^4 k}{(2 \pi)^4}
\left[\frac{k^2}{(k^2+m_1^2)(k^2+m_2^2)}\right]
\eeq
\[
= \frac{1}{16\pi^2}\left[\frac{m_2^2+m_1^2}{(m_2^2-m_1^2)^2}
+\frac{2m_1^2m_2^2}{(m_2^2-m_1^2)^3}\ln \left(\frac{m_1^2}{
m_2^2}\right)\right]\,,
\] \
\beq
I_2(m_1,m_2) = \int_E \frac{d^4 k}{(2 \pi)^4}
\left[\frac{1}{(k^2+m_1^2)^2(k^2+m_2^2)(k^2+M_W^2)}
\right]
\eeq
\[
= \frac{1}{16\pi^2}\left[\frac{m_2^2\ln(m_1^2/m_2^2)}
{(m_2^2-m_1^2)^2(m_2^2-M_W^2)}+\frac{M_W^2\ln(m_1^2/M_W^2)}
{(M_W^2-m_1^2)^2(M_W^2-m_2^2)}\right.
\]
\[
\left.
-\frac{1}{(m_1^2-m_2^2)
(m_1^2-M_W^2)}\right]\,,
\]
\beq
I_3(m_1,m_2) = \int_E \frac{d^4 k}{(2 \pi)^4}
\left[\frac{k^2}{(k^2+m_1^2)^2(k^2+m_2^2)(k^2+M_W^2)}
\right]
\eeq
\[
= \frac{1}{16\pi^2}\left[\frac{m_2^4\ln(m_2^2/m_1^2)}
{(m_2^2-M_W^2)(m_1^2-m_2^2)^2}+\frac{M_W^4\ln(M_W^2/m_1^2)}
{(M_W^2-m_2^2)(m_1^2-M_W^2)^2}\right.
\]
\[
\left.
+\frac{m_1^2}{(m_1^2-m_2^2)
(m_1^2-M_W^2)}\right]\,.
\]



\begin{thebibliography}{99}
\frenchspacing

\bibitem{etc}
S. Dimopoulos and L. Susskind, {\it Nucl. Phys.} {\bf B155}
(1979) 237; E. Eichten and K. Lane, {\ it Phys. Lett.} {\bf B90}
(1980) 125.

\bibitem{ctsm}
R. S. Chivukula and H. Georgi, {\it Phys. Lett.} {\bf B188} (1987)
99; S. Dimopoulos, H. Georgi, and S. Raby, {\it Phys. Lett.} {\bf
B127} (1983) 101; R. S. Chivukula, H. Georgi, and L. Randall,
{\it Nucl. Phys. } {\bf B292} (1987) 93.

\bibitem{bostech}
E. H. Simmons, {\it Nucl. Phys.} {\bf B312} (1989) 253;
S. Samuel, {\it Nucl. Phys.} {\bf B347} (1990) 625; A. Kagan and
S. Samuel, {\it Phys. Lett.} {\bf B252} (1990) 605 and
{\it Phys. Lett. } {\bf B270} (1991) 37.

\bibitem{carone}
C. D. Carone and E. H. Simmons, {\it Nucl. Phys.} {\bf B397}
(1993) 591.

\bibitem{colwein} S. Coleman and E. Weinberg, {\it Phys. Rev.}
{\bf D7} (1973) 1888.

\bibitem{manohar}
S. Weinberg, Physica {\bf 96A} (1979) 327;
A. Manohar and H. Georgi, {\it Nucl. Phys.} {\bf B234} (1984)
189; L. Randall, {\it Nucl. Phys.} {\bf B276} (1986) 241;
T. Applequist, Scottish Summer School (1980) 385.

\bibitem{wise}
L.F. Abbott, P. Sikivie, and M. B. Wise,
Phys. Rev. {\bf D21} (1980) 1393.

\bibitem{aleph}
D. Decamp, {\it et. al.}, Phys. Rept. 216 (1992) 253.

\bibitem{pesktak}
M.E. Peskin and T. Takeuchi, {\it Phys. Rev. Lett.} {\bf 65}
(1990) 964; {\it Phys. Rev.} {\bf D46} (1992) 381.

\bibitem{tak}
T. Takeuchi, SLAC-PUB-5730 (1992), Presented at Int. Workshop
on Electroweak Symmetry Breaking, Hiroshima, Japan, 1991.

\bibitem{coleman1}
S. Coleman, {\it Nucl. Phys.} {\bf B307} (1988) 867.

\bibitem{coleman2}
S. Coleman, {\it Nucl. Phys.} {\bf B310} (1988) 643.

\bibitem{preskill}
J. Preskill, {\it Nucl. Phys.} {\bf B323} (1989) 141.

\end{thebibliography}
\end{document}